\documentstyle[12pt,epsfig]{article}

\advance\textheight by \topskip
\begin{document}
\tolerance=100000
\thispagestyle{empty}
\setcounter{page}{0}

\newcommand{\be}{\begin{equation}}
\newcommand{\ee}{\end{equation}}
\newcommand{\br}{\begin{eqnarray}}
\newcommand{\er}{\end{eqnarray}}
\newcommand{\ba}{\begin{array}}
\newcommand{\ea}{\end{array}}
\newcommand{\bi}{\begin{itemize}}
\newcommand{\ei}{\end{itemize}}
\newcommand{\bn}{\begin{enumerate}}
\newcommand{\en}{\end{enumerate}}
\newcommand{\bc}{\begin{center}}
\newcommand{\ec}{\end{center}}
\newcommand{\ul}{\underline}
\newcommand{\ol}{\overline}
\def\epem{\ifmmode{e^+ e^-} \else{$e^+ e^-$} \fi}
\def\mathrm{\rm}
\newcommand{\WW}{$W^+ W^-$}
\newcommand{\eeww}{$e^+e^-\rightarrow W^+ W^-$}
\newcommand{\LEPONE}{LEP1}
\newcommand{\LEPTWO}{LEP2}
\newcommand{\QCD}{QCD}
\newcommand{\JETSET}{JETSET}
\newcommand{\MC}{MC}
\newcommand{\qq}{$q\bar q$}
\newcommand{\QQ}{$Q\bar Q$}
\newcommand{\qqgg}{$q\bar q gg$}
\newcommand{\qqQQ}{$q\bar q Q\bar Q$}
\newcommand{\eeWWqqQQ}{$e^+e^-\rightarrow W^+ W^-\ar q\bar q Q\bar Q$}
\newcommand{\eeqqgg}{$e^+e^-\rightarrow q\bar q gg$}
\newcommand{\eeqqQQ}{$e^+e^-\rightarrow q\bar q Q\bar Q$}
\newcommand{\eewwjjjj}{$e^+e^-\rightarrow W^+ W^-\rightarrow 4~{\rm{jet}}$}
\newcommand{\eeqqggjjjj}{$e^+e^-\rightarrow q\bar 
q gg\rightarrow 4~{\rm{jet}}$}
\newcommand{\eeqqQQjjjj}{$e^+e^-\rightarrow q\bar q Q\bar Q\rightarrow
4~{\rm{jet}}$}
\newcommand{\eejjjj}{$e^+e^-\rightarrow 4~{\rm{jet}}$}
\newcommand{\jjjj}{$4~{\rm{jet}}$}
\newcommand{\qqbar}{$q\bar q$}
\newcommand{\ar}{\rightarrow}
\newcommand{\sm}{${\cal {SM}}$}
\newcommand{\Dir}{\kern -6.4pt\Big{/}}
\newcommand{\Dirin}{\kern -10.4pt\Big{/}\kern 4.4pt}
\newcommand{\DDir}{\kern -7.6pt\Big{/}}
\newcommand{\DGir}{\kern -6.0pt\Big{/}}
\newcommand{\wwqqqq}{$W^+ W^-\ar q\bar q Q\bar Q$}

\def\Ord{\buildrel{\scriptscriptstyle <}\over{\scriptscriptstyle\sim}}
\def\OOrd{\buildrel{\scriptscriptstyle >}\over{\scriptscriptstyle\sim}}
\def\pl #1 #2 #3 {{\it Phys.~Lett.} {\bf#1} (#2) #3}
\def\np #1 #2 #3 {{\it Nucl.~Phys.} {\bf#1} (#2) #3}
\def\zp #1 #2 #3 {{\it Z.~Phys.} {\bf#1} (#2) #3}
\def\jp #1 #2 #3 {{\it J.~Phys.} {\bf#1} (#2) #3}
\def\pr #1 #2 #3 {{\it Phys.~Rev.} {\bf#1} (#2) #3}
\def\prep #1 #2 #3 {{\it Phys.~Rep.} {\bf#1} (#2) #3}
\def\prl #1 #2 #3 {{\it Phys.~Rev.~Lett.} {\bf#1} (#2) #3}
\def\mpl #1 #2 #3 {{\it Mod.~Phys.~Lett.} {\bf#1} (#2) #3}
\def\rmp #1 #2 #3 {{\it Rev. Mod. Phys.} {\bf#1} (#2) #3}
\def\cpc #1 #2 #3 {{\it Comp. Phys. Commun.} {\bf#1} (#2) #3}
\def\sjnp #1 #2 #3 {{\it Sov. J. Nucl. Phys.} {\bf#1} (#2) #3}
\def\xx #1 #2 #3 {{\bf#1}, (#2) #3}
\def\preprint{{\it preprint}}
\def\hepph #1 {{\tt hep-ph/#1}}

\begin{flushright}
{\large RAL-TR-1998-037}\\
{\large DTP/98/32}\\ 
{\large Cavendish-HEP-97/12}\\ 
{\rm August 1998\hspace*{.5 truecm}}\\ 
{\rm Revised December 1998\hspace*{.5 truecm}}\\ 
\end{flushright}

\vspace*{\fill}

\begin{center}
{\Large {\bf Spin correlations in} 
${e^+e^-\ar 4}$~jets\footnote{E-mails:  
Moretti@rl.ac.uk; W.J.Stirling@durham.ac.uk.}}\\[1.cm]
{\large 
S.~Moretti$^{a,}$\footnote{Formerly: Cavendish Laboratory, 
 Cambridge University, Madingley Road, Cambridge CB3 0HE,~UK.} 
and W.J.~Stirling$^{b,c}$}\\[0.4 cm]
{\it a) Rutherford Appleton Laboratory,}\\
{\it Chilton, Didcot, Oxon OX11 0QX, UK.}\\
{\it b) Department of Physics, University of Durham,}\\
{\it South Road, Durham DH1 3LE, UK.}\\
{\it c) Department of Mathematical Sciences, University of Durham,}\\
{\it South Road, Durham DH1 3LE, UK.}\\
\end{center}

\vspace*{\fill}

\begin{abstract}
{\normalsize
\noindent
Existence of discrepancies between experimental data and Monte Carlo predictions
for  angular distributions in four-jet production  via $e^+e^-$ 
annihilation has been known for some time at LEP1.
As such QCD processes constitute significant backgrounds to $e^+e^-\ar W^+W^-$ 
$\to$ $4$-jet 
production at LEP2, we consider the possibility that an erroneous modelling
of the helicity structure of the final-state partons
could affect the accuracy of experimental measurements of the 
 $W^\pm$ boson parameters.}
\end{abstract}

\vspace*{\fill}
\newpage

\section{Introduction and motivation}

The partonic reactions $e^+e^-\ar q\bar q gg$ and $e^+e^-\ar 
q\bar q Q\bar Q$ (see Fig.~1) constitute the 
dominant (lowest-order) component
of the annihilation rate of electrons and positrons into four jets.
They represent a benchmark process in QCD phenomenology at present
and future $e^+e^-$ colliders for several reasons. 

First, they provide a  test of the underlying $SU(N_C)$ colour 
symmetry of the strong interactions between quarks and gluons,
as the cross section is sensitive to the three fundamental colour factors 
$C_A$, $C_F$ (the Casimir operators of the fundamental 
and adjoint  representations of
the gauge group $SU(N_C)$, respectively) and $T_F$ (the normalisation of the
generators of the fundamental representation). 
These are determined
from the $SU(N_C)$ generators $(T^a)_{ij}$ and the structure constants
$f^{abc}$ via the relations
\begin{equation}\label{factors} 
\sum_{a,b}f^{abc}f^{abd*}=\delta^{cd}C_A,\quad\quad
\sum_{a}(T^aT^{a\dagger})_{ij}=\delta_{ij}C_F,\quad\quad
{\mathrm{Tr}}[T^aT^{b\dagger}]=\delta^{ab}T_F,
\end{equation}
where $a,b,..(i,j,...)$ represent gluon(quark) colour indices, yielding 
\begin{equation}\label{algebra}
C_A=N_C,\quad\quad
C_F=(N_C^2-1)/2N_C,\quad\quad
T_F=1/2.
\end{equation}
In QCD (i.e. $N_C=3$) one has $C_A=3$ and $C_F=4/3$.
The factors $C_A$, $C_F$ and $T_F$ represent the relative
strength of the couplings (squared) of the processes 
$q\rightarrow qg$, $g\rightarrow gg$
and $g\rightarrow q\bar q$, respectively (see for example Ref.~\cite{Quigg}).

Second, under the assumption that $SU(N_C\equiv3)$ is indeed the
gauge group of the theory, a measurement of these colour factors (in particular
of $T_R=N_FT_F$)
can be converted into a constraint on the value of
the number of quark flavours active at the energy scale at which the $\epem$\
annihilation takes place. For example, $N_F$ would be 
increased (by approximately 3) from its ($N_F=5$) 
Standard Model (SM) value at LEP by an additional New Physics contribution
from very light gluinos ${\mathaccent"7E g}$ 
\cite{window}
produced in the process $e^+e^-\rightarrow q\bar q
{\mathaccent"7E g}{\mathaccent"7E g}$ via a $g^*\rightarrow
{\mathaccent"7E g}{\mathaccent"7E g}$ splitting \cite{epem}.

Third, QCD four-jet production forms  a significant
background   to
$e^+e^-\ar W^+W^- \to 4$~jets  at LEP2 \cite{LEP2}, 
for both the `threshold' (total cross section)
and the `direct reconstruction' ($W^\pm $ line shape) methods 
employed in the measurement of  $M_{W}$, even after
the implementation of typical $W^\pm$ selection cuts \cite{bkgd}. 
Indeed, although biased by systematic uncertainties due to relatively
unknown `Bose-Einstein' \cite{bose} and `colour interconnection' \cite{CR}
effects, the fully hadronic signature of the two $W^\pm$s still represents 
to date  the experimentally preferred decay channel, 
because kinematic constraints can tighten the precision of the  $M_{W}$ 
measurement.  

Fourth, four-jet events represent a serious background in the
search for Higgs particles at LEP2 \cite{HiggsLEP} and 
beyond (e.g. at the Next Linear Collider \cite{ee500}) 
\cite{HiggsNLC}, both in the standard electroweak (EW) theory
and in possible extensions. In the SM, 
the dominant Higgs production channel proceeds  via the `bremsstrahlung' process
$e^+e^-\ar Z\phi$ \cite{Bjorken}, 
followed by the hadronic decays $\phi\ar b\bar b$.
In this respect, it should be recalled that $Z$ bosons have a 70\% branching
ratio into a pair of jets. Such arguments also apply to the case
of the light $h$ and heavy $H$ Higgs scalars of the Minimal Supersymmetric
Standard Model (MSSM) \cite{wrk39}. 
In addition, in the MSSM four-jet signatures arise naturally from
the decays of the $h$ scalar and the $A$ pseudoscalar produced in pairs
via $e^+e^-\ar hA$ \cite{wrk39}. 

Fifth, the four-jet channel could well be the footprint of new
{\sl sparticles} at LEP2 or beyond, 
via the production 
of two resonances both decaying hadronically. In this context, it should
be recalled that the production of several different types of supersymmetric
particles was advocated some years ago as an explanation of the apparent excess 
of four-jet events as recorded by the ALEPH Collaboration \cite{alephex}.
 For example, pairs of sneutrinos
\cite{BKP}, squarks \cite{F1} (in particular 
sbottoms \cite{GPVW}), charginos and  neutralinos \cite{GGR,DLM,CCP,F2}
 and also selectrons \cite{CGLW} can yield four-jet 
signatures. Technicolour was also suggested as a possible explanation 
of the ALEPH events \cite{techni}. 

For all of the above reasons it is clearly very important that
 four-jet events are correctly
implemented in the Monte Carlo (MC) programs which are widely used in 
phenomenological studies of hadron production at $\epem$ colliders (e.g.
HERWIG \cite{HERWIG}, JETSET/PYTHIA \cite{JETSET}, ARIADNE \cite{ARIADNE},
etc.). In this connection,  it is rather worrying that certain aspects of 
 four-jet production
are  apparently {\sl not} well described by the {\sl standard} 
`${\cal O}(\alpha_s)$ ME + parton shower (PS)' MC programs. For example, 
four-jet studies performed by the ALEPH Collaboration at LEP1 
\cite{ALEPHgluino} 
have revealed a
significant disagreement (up to $20\%$ in some regions) 
between data and MCs for distributions in the four-jet angular variables:
(i) the Bengtsson-Zerwas angle $\chi_{\mathrm{BZ}}$; 
(ii) the K\"orner-Schierholz-Willrodt angle $\Phi_{\mathrm{KSW}}$;
(iii) the (modified) Nachtmann-Reiter angle $\theta_{\mathrm{NR}}^*$;
(iv) the angle between the two least energetic jets $\theta_{34}$.
This is well illustrated by Fig.~2 of Ref.~\cite{ALEPHgluino}.

The above angular quantities are defined in terms of the four-momenta of the
particles in the final state and  are particularly sensitive to the 
 flavour (fermionic and bosonic) composition  of four-jet events, 
in the sense that the different spins carried by quarks and gluons 
induce very different relative orientations between the two planes defined  
by the directions of the {\sl primary} (i.e. from the 
$\gamma^*,Z^{(*)}$-splitting)
and {\sl secondary} (i.e. from the virtual gluon splitting into quarks or 
gluons) parton pairs\footnote{To be more precise, the
angle $\theta_{34}$ was originally introduced to discriminate between the
double bremsstrahlung  (Fig.~1a--b) and 
triple-gluon-vertex (Fig.~1c) diagrams of the \qqgg\  subprocess 
(see Ref.~\cite{t34}), in order
to provide direct proof of the non-Abelian structure of QCD. In fact, the
variable $\theta_{34}$ is more sensitive to the gluon propagator singularity
in the graphs of Fig.~1c, which is of course absent in those of Fig.~1a--b
(see also the discussion in Sect.~3 below), 
than it is to spin correlations among partons.}. 
For this reason, the angles $\chi_{\mathrm{BZ}}$,
$\Phi_{\mathrm{KSW}}$, $\theta_{\mathrm{NR}}^*$ and $\theta_{34}$
have profitably been  used to fit the theoretical predictions to the data
in order to measure the three colour factors $C_A$, $C_F$ and $T_F$.
It can therefore be argued that the disagreement might be
evidence that the  MCs do not  provide a correct description of
the spin correlations among  the various partons over the all phase space.
 Conversely, ${\cal O}(\alpha_s^2)$ ME programs
(for example, the  `${\cal O}(\alpha_s^2)$ ME 
+ $\mbox{string fragmentation}$ model' as implemented in JETSET, i.e.
with no parton shower) yield a much
better angular description of four-jet final states (see Fig.~3 of 
\cite{ALEPHgluino} and \cite{GC}).
Indeed, all spin  correlations are naturally included in a full
matrix element calculation (perturbative QCD predicts very specific 
orientations among 
the four final-state partons,  see for example Ref.~\cite{QCDbook}), 
but they are not necessarily present  in a 
PS emulation of the four-jet final state, though their availability
(in the infrared limit, where soft and collinear correlations can
be factorised in analytic form) is a feature of some of the 
above mentioned MC codes (for example, for the HERWIG implementation, 
see Refs.~\cite{Ian1,Ian2}).

It might appear, therefore, that ${\cal O}(\alpha_s^2)$ ME programs
are the preferred tool for analysing QCD four-jet production at LEP (and
beyond).  However, the problem here is  that  
such ME models contain `ad-hoc' hadronisation which  is adjusted
to give a good description of some LEP1 data \cite{ALEPHgluino}
(see also Ref.~\cite{OP}) but cannot be reliably 
extrapolated to  higher (e.g. LEP2) energies. Furthermore, it is a well
known fact that their description of the sub-jet structure is very poor
(see Ref.~\cite{Andre} for an overview).

The deficiencies of these various approaches have been known for some time.
Indeed, a general concern about, on the one hand, the inability of the 
standard MC programs to reproduce accurately
four-jet angular quantities typical of 
parton level QCD and,
on the other hand, the limitations of the ME implementation in quantifying
hadronic multiplicities, already emerged at the time of the CERN LEP2 Workshop, 
see Ref.~\cite{iantor}. In this respect, a key point not addressed there 
and that in our opinion deserves urgent
attention is to assess whether or not also the angular 
behaviour of four-jet QCD events that survive $W^+W^-$ selection criteria
is poorly modelled by the standard parton showers, as well 
as to quantify the sort of 
differences that one should expect from the two approaches.
While it is the purpose of this study to specifically address this question, we
also attempt here to understand the source of the disagreement between the PS 
and the ME predictions for the above angular variables. 

The plan of the paper
is as follows. In the next Section we describe our method of computing
four-jet rates  using perturbative QCD. In Section 3 
we present our results, and Section 4 contains a brief summary
and some conclusions. 

\section{$e^+e^-\ar q\bar q gg$ and $e^+e^-\ar 
q\bar q Q\bar Q$ events}

In order to generate the exact leading-order (LO) 
QCD predictions for the four-parton
processes we use the {\tt FORTRAN} codes already exploited in 
Refs.~\cite{noiPL,noiNP,noiPr}. Such programs include not only all the angular 
correlations between primary and secondary partons,
but also all quark masses and both the $\gamma^*$ and $Z^{(*)}$ 
intermediate contributions.
As numerical inputs for the electroweak parameters we take
$\alpha_{em}= 1/128$ and  $\sin^2\theta_W=0.23$, while
for the $Z$ boson mass and width we use
$M_{Z}=91.1$~GeV and $\Gamma_{Z}=2.5$~GeV, respectively.
For the $W^\pm$ boson, we use $M_{W}=80.23$~GeV and 
$\Gamma_{W^\pm}=2.08$~GeV.
The $u$, $d$ and $s$ quarks are taken to be massless,
while for $c$ and $b$ we take $m_c=1.35$~GeV and $m_b=4.95$~GeV.
The strong coupling constant $\alpha_s$ has been computed at two-loop
order (NLO), for $N_F=5$ active flavours, 
with $\Lambda_{\mathrm{QCD}}=190$~MeV and scale choice $\mu=\sqrt s$,
where $\sqrt s$ is the centre-of-mass (CM) energy.

In order to define a four-jet sample we need to introduce a jet algorithm.
In the present study we will use the so-called `Durham' (D) one\footnote{See 
Ref.~\cite{schemes} 
for a survey of jet clustering algorithms and their properties.} \cite{DURHAM},
which is widely used at LEP and SLD, 
based on the `measure'
\begin{equation}\label{DURHAM}
y^D_{ij} = {{2\min (E^2_i, E^2_j)(1-\cos\theta_{ij})}
\over{s}}.
\end{equation}

In the above equation $E_i$ and $E_j$ are the energies
of the particles $i$ and $j$, and $\theta_{ij}$ is their relative
angle (with $i\ne j=1,\dots 4$). In our tree-level ME calculations, the 
four-jet cross section for a given algorithm is simply equal to
the four-parton cross section with a cut $y_{ij}\ge y_{\mathrm{cut}}$ on all
possible parton pairs $(ij)$. In contrast, in our parton-shower studies, 
we need to define a jet clustering procedure as well. In these cases, 
we have adopted 
the so-called E-scheme, so that the four-momentum of a new cluster (or
pseudoparticle) $k$ is found from its constituents $i$ and $j$ by simple 
addition,  $p_k = p_i + p_j$.
The joining procedure is repeated until all pairs of clusters have a
separation above $y_{\mathrm{cut}}$. This final set of clusters is
called jets.

At this point, we should further mention that we have also tried out 
three other jet clustering schemes, such as the `Jade' \cite{JADE},
`Geneva' \cite{GENEVA} and `Cambridge' \cite{CAMBRIDGE} ones. Indeed,
we have verified that none of our conclusions  
depends on the details of the jet algorithm used.
 
For completeness, we recall the definitions of the four angular variables
\cite{BZ,KSW,NR,NRmod} introduced in the previous Section. 
One first orders the jets 
in energy such that ${E_1\ge E_2\ge E_3\ge E_4}$. This way,
one can identify the two most energetic jets with those originated
by the quarks produced in the $Z$ decay, and the other two with those
produced in the $g^*$ splitting (although such a procedure can be affected
by systematic errors due to possible mis-assignments, see 
\cite{nonAbel,ioebas}).
Then, in terms of the three-momenta
$\vec{p}_1,\dots {\vec p}_4$ of the energy-ordered 
jets $1,\ldots 4$, the angles
$\theta_{\mathrm{NR}}^*$, 
$\chi_{\mathrm{BZ}}$ and $\theta_{34}$ are defined as 
\be
\label{def:NR}
\theta_{\mathrm{NR}}^*=\angle({\vec p}_1-{\vec p}_2,{\vec p}_3-{\vec p}_4),
\ee
\be
\label{def:BZ}
\chi_{\mathrm{BZ}}=
\angle({\vec p}_1 \times {\vec p}_2,{\vec p}_3 \times {\vec p}_4),
\ee
and
\be
\label{def:34}
\theta_{34}=\angle({\vec p}_3,{\vec p}_4).
\ee
For $\Phi_{\mathrm{KSW}}$ 
we actually use the `modified' definition proposed in
Ref.~\cite{ioebas} (hereafter denoted by $\Phi_{\mathrm{KSW}}^*$), 
which is 
more sensitive than the original to the flavour composition (see 
Refs.~\cite{KSW,ioebas}). Thus, in  events for which 
\be
\label{def:KSW1}
|{\vec p}_1+{\vec p}_3| > |{\vec p}_1+{\vec p}_4|
\ee
we define
\be
\label{def:KSW2}
\Phi_{\mathrm{KSW}}^*=
\angle({\vec p}_1\times{\vec p}_3,{\vec p}_2\times{\vec p}_4),
\ee
whereas in the opposite case
 we define $\Phi_{\mathrm{KSW}}^*$ with ${\vec p}_3$ and
${\vec p}_4$ interchanged\footnote{Note that the definition given in 
Eqs.~(\ref{def:KSW1})--(\ref{def:KSW2}) is equivalent to the
original $\Phi_{\mathrm{KSW}}$ angle 
\cite{KSW} in events where the thrust
axis is directed along ${\vec p}_1+{\vec p}_3$ or ${\vec p}_1+{\vec p}_4$.}.

\section{Results}

Our results are presented in Figs.~2--8 and in Tab.~I. They include
no Initial State Radiation (ISR) at any of the energies, neither in the ME nor
in the PS calculations. This is done in order to simplify the discussion, 
as we have verified that the relative behaviours
of the two implementations are unaffected by the inclusion of such emission.
In addition, notice that in the course of our analysis we will at times 
modify the value of the 
jet resolution parameter $y_{\mathrm{cut}}$ adopted to define the four-jet
sample. (In particular, the algorithm and the resolution used
in Fig.~3 are the same as those of
the ALEPH study \cite{ALEPHgluino}, enabling us to
make quantitative comparisons of our findings with the results given there.)
This will help to illustrate the generality of the phenomenology that 
we will describe and, in particular, will help to distinguish properties
of the underlying matrix elements from artifacts of the jet definition.

In the Introduction we argued that the disagreement between data
and MCs could be due to a lack of adequate spin 
correlations among the partons generated in the phenomenological programs.
In order to illustrate the sensitivity of the four angular variables 
to the helicity structure of the $\gamma^*,Z^{(*)}$ decay products we plot
in Fig.~2 the distributions in $\chi_{\mathrm{BZ}}$, 
$\Phi_{\mathrm{KSW}}^*$,
$\theta_{\mathrm{NR}}^*$ and  $\theta_{34}$
for the following subprocesses: (i) the four-quark component as obtained
from the exact ME (tree-level) calculation;
(ii) the same in a PS-like model obtained by averaging over the helicities
of the virtual gluon in Fig.~1d (which removes the spin correlations
between the two quark pairs in the final state); 
(iii) the triple-gluon-vertex component
of two-quark-two-gluon events from the QCD ME (diagrams in Fig.~1c);
(iv) the same in a PS-like model again simulated by averaging over the 
virtual gluon spins. Note that we have not presented similar distributions
for the case of the double bremsstrahlung component of the 
two-quark-two-gluon ME (see Fig.~1a--b), for two 
reasons\footnote{See Ref.~\cite{ioebas} for some typical
distributions in the above angles for all the components of the
$e^+e^-\ar 4$-parton process.}. First,
it is strongly suppressed (by roughly a factor $C_A/C_F=9/4$ \cite{QCDbook}) 
with respect to the triple-gluon-vertex contribution. Second, 
it is dominated by two infrared (i.e. soft and collinear) 
splittings of the virtual quarks into quark-gluon pairs, a dynamics
 which is well described 
by the PS implementation of the MC programs (as evidenced by their
success in describing the phenomenology of $e^+e^-\ar q\bar qg$ production 
\cite{QCDbook}). In general,  its angular behaviour
is similar to that of the non-Abelian diagrams.

We emphasise that the results displayed in
 Fig.~2 (and also Fig.~4 below)  should be regarded  more as a
useful exercise for understanding the underlying 
angular correlation properties of four-parton events in a `toy model', 
rather than as a quantitatively reliable estimate of the relative behaviour of 
the exact QCD matrix elements  vs. the MC parton shower description.
In fact, two aspects should be noticed. First, the procedure of
rewriting the diagrams in Fig.~1c and d as the product of the ME
for $e^+e^-\ar q\bar q g^*$ times those of the decays $g^*\ar gg$ and 
$g^*\ar Q\bar Q$, respectively, and of averaging
over the spins of the intermediate state (so that 
the secondary $g^*\ar gg$ and $g^*\to Q \bar Q$
splittings are azimuthally symmetric about the virtual gluon direction), 
differs in several respects from
that implemented in the MC programs. For example, here the large-angle 
splitting is correctly reproduced whereas 
in MC programs only the collinear part is correctly represented.
In contrast, in our example, the soft splitting is 
described only at lowest
order whereas in the parton cascade higher order (logarithmically) 
enhanced terms are  also included. Furthermore, as already mentioned in the
Introduction, some of the most sophisticated MC programs do include azimuthal 
and polar correlations in the parton branching (in the soft and collinear
limit), so that 
our toy model of PS should be further regarded as an extreme condition,  
and for this reason particularly useful for our purposes. 
Secondly, our procedure of separating the Abelian and 
non-Abelian components 
in the two-quark-two-gluon partonic MEs by simply
retaining the amplitudes associated with Fig.~1a--b and 1c respectively,
and neglecting their colour structure,  is clearly non-gauge-invariant. 
However, we believe  our distributions  approximate quite accurately 
the true phenomenology of the four angular variables.
In the first case, we do not expect the inclusion of large
angle dynamics nor the neglect of  
logarithmically enhanced soft radiation to
modify significantly
 the pattern of the angular correlations in the virtual gluon splitting.
In the second respect, we stress that we have normalised all distributions
to unity, since gauge invariance is more likely to affect the
overall rates of the various components we have separated, rather than
their angular shapes.
We are  therefore confident that the distributions shown
 in Fig.~2 (and 4)   are an accurate representation of the 
behaviour produced by the corresponding diagrams in the full amplitude squared,
and we can certainly use them as a guide to eventually pin-pointing the
source of the discrepancies revealed in Ref.~\cite{ALEPHgluino}.

By looking at the $\chi_{\mathrm{BZ}}$  distribution in Fig.~2, we see a clear 
peaking of the \qqQQ\ component of the QCD ME  
 around $90^\circ$,
indicating the preference for the plane of the  two secondary quark jets to be
orthogonal to the plane of the two  primary  ones. In contrast, the two
secondary gluons from the triple-gluon vertex component 
prefer to be produced 
in the plane of the primary quark-antiquark pair
(see also Fig.~9 of Ref.~\cite{wrksp}). If in our toy model
we switch off the angular correlations  for  the \qq\QQ\ final state 
we obtain a distribution 
that is significantly flatter, as expected.
The `decorrelated' version of the \qqgg\ matrix element 
only including the triple-gluon-vertex diagrams 
gives a distribution similar to that of the decorrelated \qq\QQ\ model.
The situation for the other three angular variables
 presented in Fig.~2, $\Phi_{\mathrm{KSW}}^*$, 
$\theta_{\mathrm{NR}}^*$ and $\theta_{34}$, is very much
the same as for $\chi_{\mathrm{BZ}}$. That is,
the two decorrelated versions of the QCD MEs are always significantly
different from the `correct' QCD ME predictions\footnote{Note that
in Fig.~2
of Ref~\cite{ALEPHgluino}  the distributions in the Bengtsson-Zerwas
and (modified) Nachtmann-Reiter angles are `symmetrised' by plotting
with respect to
$|\cos\chi_{\mathrm{BZ}}|$ and $|\cos\theta_{\mathrm{NR}}^*|$, respectively.
This is because there exists an arbitrariness
in the choice of the sign of the cosine of these angles. In contrast, 
 in our Figs.~2 and 4 we have plotted 
the full angular range of the two variables, in order to gain more 
insight into the underlying problem of the parton shower implementation. Our
 distributions are not symmetric in themselves, since  energy
ordering of the jets is used (see for example
the discussion in Ref.~\cite{ioebas}).}.

Therefore, Fig.~2 clearly confirms the
usefulness of the four angles in testing the underlying structure of the QCD 
processes in the presence of the full ME spin correlations. Conversely,
if the described spin dynamics is not taken into account exactly,
then their discriminating power between \qqQQ\ and \qqgg\ events could 
be significantly reduced. This should be especially borne in mind when
contemplating high-precision QCD analyses such as those aiming to determine
the $C_A$, $C_F$ and $T_F$ colour factors (and/or the possible
existence of light gluinos).

For completeness, and in order to  justify the confidence we have in 
our parton-level results,
we also present in Fig. 3 the ratios between our ME and the PYTHIA 
results for the
four angular distributions
as obtained by adopting the same  jet algorithm
and resolution used in Ref.~\cite{ALEPHgluino}. Here, in the MC implementation,
partons are clustered before hadronisation until four jets are left. 
In addition, among all possible configurations of
the latter, we only retain those for which the minimum of the $y_{ij}$ measures
is greater than $y_{\mathrm{cut}}$. 
The aim is  to reproduce here the salient features
of Fig.~2 of Ref.~\cite{ALEPHgluino}, our ME outputs mimicking the data of that
experimental study. In this respect, note that in Fig.~2 of 
Ref.~\cite{ALEPHgluino} a full simulation (DYMU + JETSET + ALEPH
detector) was performed and the angular variables were
computed at hadron  level. However, 
given the good agreement between the two figures, we believe
that the main conclusions that we will draw
from the forthcoming studies are unaffected by our simplification.
We will now set aside further considerations of  
the subject of precision QCD analyses and move on to another aspect of 
four-jet phenomenology.

We consider here the more topical issue of $W^+W^-$ production and 
hadronic decays at LEP2.
We have already mentioned that the QCD processes $e^+e^-
\to \gamma^*,Z^{(*)} \to$ \qqgg, \qqQQ\  constitute significant 
backgrounds to
\WW\ $\to$ $4$-jet production at \LEPTWO, up to $20\%$ of the signal
depending on the event selection criteria used. 
Thus, it is crucial that the MC generators
used in the experimental studies are able to correctly describe the salient
features (that is, the overall rate and the differential distributions)
of the  QCD four-jet final states that are compatible with
the $W^+W^-\to4$-jet kinematics.

A preliminary study of this problem was presented in Ref.~\cite{wrksp}.
In particular,
it was investigated there whether \QCD\ events which pass the \WW\ event 
selection populate
the ranges of the four-jet angular variables where the MCs have been
shown to have problems describing the \LEPONE\ 
data. Fig.~8 of Ref.~\cite{wrksp},
obtained from simulations of  QCD and $W^+W^-$ events
at 172~GeV binned according to the four angular variables listed above,
seemed to suggest that this is indeed the case.
For example, using rather
generic $W^+W^-$ selection cuts, it was shown  that  the 
$W^+W^-$ signal and the
 \QCD\ background   do cluster in  the same angular regions. 
In particular, from Fig.~2 of Ref.~\cite{ALEPHgluino}
and Fig.~8 of Ref.~\cite{wrksp},
one can identify the following ranges of concern
\begin{equation}\label{concern}
\vert \cos\chi_{\mathrm{BZ}} \vert , \vert\cos\theta^*_{\mathrm{NR}}
  \vert  < 0.7,
\qquad
\vert  \cos\Phi^*_{\mathrm{KSW}}\vert , \vert \cos\theta_{34} \vert  <  0.5 .
\end{equation}
Using the \QCD\ predictions  taken from PYTHIA and the 
\WW\ predictions  from  KORALW \cite{koralw}, one finds that
at LEP2 approximately half the \QCD\ events have, for example, 
$  \vert \cos\theta_{34} \vert  <  0.5  $,
a  region largely populated by the signal and
where the JETSET PS fails to describe the \LEPONE\ QCD data  by up
to 15\% \cite{GC} ! 

It is therefore  extremely important to investigate further such disagreement,
particularly at LEP2 energies. For example, it is not impossible that
 $W^+W^-$ selection cuts might somehow affect the relative
performances of the ME and PS implementations in describing the 
spin correlations. In particular, the constraints
devised to disentangle the $W^+W^-$ signal at LEP2 could be such that
the decorrelated MEs (and therefore 
 also the MCs that they are supposed
to emulate) approximate the exact results
of the QCD MEs   at LEP2 to a  much better extent than they do at LEP1, 
thus reducing the severity of the problem. 

Unfortunately, this possibility does not   appear to be true,
as can be deduced from  Fig.~4. This shows the  same 
distributions as Fig.~2, but now at 
 $\sqrt s=172$~GeV.
In addition to the processes studied at $\sqrt s=M_Z$
we now include the  corresponding angular distributions 
obtained at parton level from the 
\eeWWqqQQ\ ME  introduced and described in Ref.~\cite{CRpap}.
As a `typical'  \WW\ selection cut we have required 
$|M_{ij}-M_{W}|<10$ GeV for at least two pairs of partons $(ij)$.
We see from Fig.~4  that, in the regions populated
by $W^+W^-$ events, the differences
in the shapes of the ME and PS-like implementations are still significant
for $\chi_{\mathrm{BZ}}$ and $\Phi_{\mathrm{KSW}}^*$,
somewhat reduced for $\theta_{\mathrm{NR}}^*$, and
almost non-existent for
$\theta_{34}$. However,
before claiming that $\theta_{34}$ has become a `safe' quantity at LEP2, 
we should  recall that, unlike the other variables,
the angle between the two least energetic jets
(generally, those produced in the virtual gluon splitting or by 
double gluon bremsstrahlung)
is  strongly dependent on the choice of the jet algorithm and/or the
resolution parameter $y_{\mathrm{cut}}$ (i.e. on the higher-order
terms of the perturbative expansion \cite{a3} not included here).
As we have already mentioned, it is a direct measure 
of the singularity which appears when the two secondary partons
become collinear. (This singularity is of course absent
for the  $W^+W^-$ distribution, hence its flatness.) In other words, 
$\theta_{34}$ is a rather `infrared unstable'
variable that, in our opinion, is not a reliable indicator of the
underlying properties of the QCD four-jet matrix elements.

Finally we turn to a comparison of the angular distributions at LEP2
obtained from the ${\cal O}(\alpha_s^2)$ QCD
MEs and from two of the most widely used MC event generators, the up-to-date
HERWIG 5.9 and PYTHIA 6.1, both of these using the
`${\cal O}(\alpha_s)$ ME + PS' approximation. We require them 
to reconstruct exactly four jets 
from the parton level before hadronisation, using the 
Durham jet finder with $y_{\mathrm{cut}}=0.002$.
Fig.~5a shows the distributions in the four angular variables
at the LEP2 energy of 172~GeV with the usual $W^+W^-$ selection
cut $|M_{ij}-M_{W}|<10$ GeV enforced. 

Even assuming that the overall normalisation is the same in all cases
(the distributions in Fig.~5a are all normalised to unity),
we see clear discrepancies between the ME (solid histogram) and 
PS (dashed and dotted histograms) predictions, although apparently 
 less pronounced than at LEP1 
(see Refs.~\cite{ALEPHgluino,GC}).  Furthermore, the discrepancies
are significant in the regions where
the $W^+W^-$ signal peaks, especially for  
 the Bengtsson-Zerwas and the (modified)
K\"orner-Schierholz-Willrodt angles, well in line with our previous 
findings in the toy model. 
It is also evident that a simple rescaling of the distributions
in the $W^+W^-$ populated regions does not resolve the differences.
To quantify the
size of the effects seen in Fig.~5a, we reproduce in Tab.~I (first line)
the fraction of events found in various angular intervals
(typically, around the maximum of the $W^+W^-$  distributions) for the ME
predictions and those from HERWIG and PYTHIA. For reference, we also 
include the event fractions for the $W^+W^-$ 
signal. From Tab.~I we see that the  differences 
between the QCD ME and the PS predictions can be up to about
10--15\%\footnote{We
have verified that the same size of effects are obtained using
other jet algorithms and resolution parameters.}.
In addition, HERWIG and PYTHIA behave rather similarly.
Thus, we conclude that one should expect significant discrepancies
in the shapes of the angular distributions predicted by the ME and PS
models when performing  $W^+W^-\ar$ 4-jet analyses. 

 It is more difficult to draw any firm conclusions
about possible differences in overall normalisation.
The tree-level ${\cal O}(\alpha_s^2)$ ME four-jet {\sl rate} is rather unstable
against the
effects of NLO QCD corrections (see Ref.~\cite{a3}), with
$K$-factors which can be of ${\cal O}(100\%)$.
However, the {\sl shape} of the angular 
distributions appears to be  much more stable, implicitly demonstrating the
relevance of the non-infrared dynamics in determining the behaviour of the
four-jet angles. For example at LEP1, in the Durham algorithm 
and for $y=0.008$
(the same set-up as in Ref.~\cite{ALEPHgluino}), the typical $K$-factor of
the angular
spectra  normalised to unity always lies in the range $(1\pm0.025)$
for all angular variables, see Figs.~1--4 of Ref.~\cite{anglesNLO}.
Therefore, NLO effects alone cannot account for the discrepancies seen
in Fig.~2 of Ref.~\cite{ALEPHgluino}, nor do we expect them to be the reason
for the differences noted in Fig.~5a here.
For reference, we note that the JETSET
LEP1 four-jet rates of \cite{ALEPHgluino} and \cite{GC} are obtained 
by using an acceptance-rejection algorithm such that
the three-jet rate of the generated events matches that given by
the ${\cal O}(\alpha_s)$ ME. 

As we have mentioned the interplay between LO and NLO results in ME
calculations, we would like to make one further comment.
So far, we have compared the MEs at LO against the PS approximations
of the MC programs, the latter defined by selecting exactly $n=4$ jets
with $y_{ij}^{\mathrm{min}}>y_{\mathrm{cut}}$ (with $i\ne j=1, ... 4$). 
Indeed, an alternative procedure which can be adopted to perform the 
comparison is to
include also PS events with $n\ge4$ jets (above the resolution) and eventually
cluster these into exactly $n=4$ jets. In fact, 
although the difference between the two approaches is next-to-leading
and higher orders (recall that our ME are LO only), we have just mentioned
that NLO effects do not alter the lowest order shapes of the MEs significantly
(so should also be the case for NNLO, etc.). Furthermore, it is well known
that in many cases summing over all possible radiation inclusively
gives smaller ME/PS corrections than does vetoing resolvable radiation.
Therefore, we present in Fig.~5b the usual angular distributions, now calculated
from the two MCs in the new fashion. Indeed, 
the differences between the
ME and the PS results are still sizable, as can be seen by 
looking at the second lines of Tab.~I.

Therefore, at this point it seems clear that a PS implementation 
based on  ${\cal O}(\alpha_s^2)$ QCD calculations for the hard scattering
is needed in order to perform studies of four-jet events such as those outlined
so far. In this respect, we note that an
`${\cal O}(\alpha_s^2)$ ME + parton shower + $\mbox{cluster fragmentation}$' 
option, based on both the correct second-order ME dynamics (from Refs.
\cite{ERT,Giele}) of the partons and 
supplemented with the appropriate showering of the latter, 
thus avoiding the  shortcomings of the 
`${\cal O}(\alpha_s^2)$ ME + $\mbox{fragmentation}$' (i.e., without PS)
approach, will soon be publicly available in the new HERWIG version 6.1 \cite{herwig61}\footnote{Some 
progress in the same direction as in HERWIG is being made also
in the context of the PYTHIA environment \cite{Andre}.}. 
The dashed curves in Fig.~3, showing the ratios of our 
`${\cal O}(\alpha_s^2)$ MEs' to the HERWIG 6.1 
`${\cal O}(\alpha_s^2)$ ME + PS' implementation, at the parton level, well 
illustrates this improvement, allowing for residual differences between the
two rates due to possible mis-assignments of the jet algorithm
reconstructing the HERWIG parton shower backwards to the original four-parton
state. The improvement is very much evident in the case of the angles
$\chi_{\mathrm{BZ}}$, $\Phi_{\mathrm{KSW}}^*$ and  
$\theta_{\mathrm{NR}}^*$. The only exception occurs for $\theta_{34}$,
where differences between the two approaches can still be relevant, 
especially when $\cos\theta_{34}\ar1$.
However, this is not surprising, as we have already highlighted the 
sensitivity of this variable to higher-order effects in the
collinear limit, which are indeed embodied in the PS evolution but not
in the LO MEs (nor in the `${\cal O}(\alpha_s^2)$ ME + hadronisation' model).
Not surprisingly,  the full NLO corrections to the shape of this angular
distribution
can be as large as 15\%, again in the region $\cos\theta_{34}\ar1$, for example
in the Cambridge scheme \cite{anglesNLO}. The appropriate treatment of
such Sudakov effects due to soft-gluon emission \cite{QCDbook} are beyond the 
scope of the present study 
and will be addressed elsewhere \cite{herwig61}.

One might now wonder whether the above parton-level behaviours
can survive the hadronisation 
process. We address this in  Fig.~6, where the usual four angular distributions
are plotted, but now at the {\it hadron} level (including 
both charged and neutral
tracks in the jet reconstruction and allowing for the decays of the heavy
hadrons), as obtained by HERWIG 6.1, using both 
${\cal O}(\alpha_s)$ and ${\cal O}(\alpha_s^2)$ initiated PSs.
From Fig.~6, it is evident that the soft dynamics of the hadronisation stage
does not remove the partonic differences between the two approaches. 
Furthermore, by comparing
Fig.~5a to Fig.~6, one can see that the discrepancies
are of the same sort at the two QCD stages, being possibly
larger at the hadron level in the case of the Bengtsson-Zerwas angle.
Therefore, it is more than plausible that `partonic' angular effects
are the source of the differences between the current PS MCs and the LEP1 data.

A  very relevant issue which can be addressed at this point is
whether such effects can influence the determination of physical parameters
whose measurement depends on the estimation of the size and topology of the 
four-jet background from QCD. One  important and topical example concerns 
the measurement of the $W$ mass at LEP2, reconstructed from
fully hadronic decays of $W^\pm$ pairs, as discussed in the Introduction.
However, in order to give quantitatively reliable estimates in this case, one 
should take into account various other non-trivial effects such as 
initial-state QED radiation, 
neutrinos that escape without detection, cracks in the detector acceptance,
experimental resolution and efficiency, etc. 
In addition, we note that the 
various selection procedures for $W^+W^-\ar$ 4-jet candidate
events differ significantly from experiment to experiment.
Therefore it is clear that a {\it full} study of the effect on the $W$ mass can
only be carried out within the context of a complete
detector simulation, which is beyond the scope of this paper.

Nevertheless, we can address this issue here at a rather more modest level,
to gauge the size of the effect.
Thus, we run  HERWIG once more  to produce 
hadronic final states via the signal process $e^+e^-\ar W^+W^-\ar
q\bar q Q\bar Q$  ({\tt IPROC=200}, as in 5.9) and the
background
$e^+e^-\ar \gamma^*,Z^*\ar q\bar qgg,q\bar q Q\bar Q$, the latter generated
both via ${\cal O}(\alpha_s)$ ({\tt IPROC=100}, as in 5.9) and 
${\cal O}(\alpha_s^2)$ ({\tt IPROC=600}, new to 6.1) MEs. Again,
we require exactly four jets to be reconstructed 
by means of the Durham scheme with 
$y_{\mathrm{cut}}>0.002$ and such that 
$|M_{ij}-M_{W}|<10$ GeV for at least two pairs of jets $(ij)$, but this time 
only within the angular regions typically populated by the
signal. Following Fig.~5a, we have chosen
the following constraints\footnote{Given the remark in Footnote 7,
recall that
we ought to symmetrize the cuts in the Bengtsson-Zerwas and Nachtmann-Reiter
angles.}:
\begin{equation}\label{cuts}
|\cos\chi_{BZ}|>0.5,
\qquad
\cos\Phi_{\mathrm{KSW}}^*<-0.5,
\qquad
|\cos\theta_{\mathrm{NR}}^*|>0.5  
\qquad
|\cos\theta_{34}|<0.8.
\end{equation}
Next, we plot an `average' $W^\pm$ mass, hereafter denoted by 
$M_{\mathrm{ave}}$, for each event generated.
 There are several reasons for choosing the  average rather than 
 the two individual masses. For example,
 the mis-assignment of one particle from one $W^\pm$ to
the other reduces the first mass and increases the second, leaving the
average less affected than each separately. In fact in  each event
 there are
three possible jet pairings, each giving one potential average $W$
mass.  Of these three, we exclude the one where the two most energetic
jets are paired with each other, since kinematically this is seldom
the correct combination. The  plot in Fig.~7 shows the shape
of such a distribution 
for the three cases {\tt IPROC=200} (solid line),
{\tt IPROC=100} (dashed line) and
{\tt IPROC=600} (dotted line). From this figure
the different shape of the two descriptions of the backgrounds is rather
clear, particularly
in the vicinity of the $W^\pm$ mass resonance.

As a further step in our study we bin (see upper
part of  Fig.~8) the average $W^\pm$ mass spectra
as obtained by summing the signal and background rates, each with the
appropriate normalisation as computed by HERWIG,
in the two possible cases, depending on whether
the ${\cal O}(\alpha_s)$ (dashed line)
or the ${\cal O}(\alpha_s^2)$ (dotted line)  MEs are
used in the MC event generator to simulate the QCD noise. The ratio in
each bin between the latter two distributions is shown in the lower
plot of Fig.~8 (solid line). Indeed,  the total spectra  also differ
significantly  and the ratio between the two is not constant around 
$M_W$.

To investigate the possible consequences of the results shown in Figs.~7--8
 on the
determination of the $W^\pm$ mass, we perform a {\tt MINUIT} \cite{minuit}
fit on the two total distributions, with a fitting function of the form
\begin{equation}\label{fitf}
f(m)=c_1\frac{c_2^2 c_3^2}{(m^2-c_2^2)^2+c_2^2 c_3^2}+g(m)
\end{equation}
where the term $g(m)$ is meant to simulate a smooth background
due to mis-assigned jets induced by the clustering algorithm.
For the latter, we adopt three different possible choices
\begin{eqnarray}\label{fitg}
g(m) = \left \{ \begin{array}{c}  0, \\[3mm]
                                  c_4+c_5~(m-c_2)+c_6~(m-c_2)^2, \\[3mm]
                                  c_4\frac{1}{1+{\mathrm{exp}}((m-c_5)/c_6)},
\end{array} \right. 
\end{eqnarray}
that is, a null, a three-term polynomial and a smeared
step function (motivated by the kinematical-limit shoulder at large masses).
Note that in eq.~(\ref{fitf}) we have assumed a Breit-Wigner shape 
characterised by a peak height $c_1$, a position $c_2$ and a width $c_3$,
corresponding to the normalisation, $M_W$ and $\Gamma_W$, respectively, 
of the two distributions in Fig.~8. 

To first approximation, then, the difference between the two values
of the coefficient $c_2$ as obtained from fitting the two curves 
is a measure of the typical
size of the systematic error that could be introduced in the experimental 
measurement of the $W$ boson mass by a mis-modeling of the QCD four-jet
background. We find this value to be around 10~MeV if we neglect
altogether the intrinsic mis-assignment background or if we assume for the
latter a step function, whereas the difference somewhat decreases for
a polynomial background, down to 2 MeV. Though not dramatically large, 
such numbers are nonetheless comparable to the expected size of the final 
systematic error on $M_W$ from all other possible sources (as estimated 
for example at
the LEP2 Workshop \cite{LEP2}) of about 50~MeV.

\section{Summary and conclusions}

In this paper we have investigated discrepancies
in the predictions for distributions of typical four-jet angular variables
obtained from  matrix element and parton shower calculations. 
Comparison of the latter with LEP1 data from ALEPH revealed differences
of up to 20\%, while the ME calculations are in good agreement with 
the same sample. 
By constructing toy parton-shower models, using matrix elements
in which the spin correlations
between the secondary (i.e. from the $g^*$ splitting) $Q\bar Q$ and $gg$ 
pairs and the primary
(i.e. from the $\gamma^*,Z^{(*)}$ decay) $q \bar q$ ones are switched off,
we concluded that the source of the disagreement 
resides in the limited capability of the MC parton shower of reproducing
the exact partonic spin correlations away from the infrared limit,
if the `${\cal O}(\alpha_s)$ ME' is used to generate the hard QCD scattering.

However, the main concern of our analysis was the impact that similar
effects could have on four-jet studies at LEP2, in the context of
$W^+W^-\ar4$-jet phenomenology and $W^\pm$ mass determinations.
The preliminary exercises that we have carried
out in this paper  show that the shape of the angular
distributions is not reproduced accurately by the MC parton showers at LEP2
either. Furthermore, these discrepancies
survive even {after} the $W^+W^-$ selection criteria are implemented
as well as the hadronisation phenomenon.

It is possible, therefore, that at present the contribution of the QCD 
four-jet background is {\it not} being correctly
simulated in candidate $W^+W^-$ samples,
and this could constitute a non-negligible source of error
in the $M_W$ determination at LEP2. 
We regard this as the most important conclusion of our paper, 
based on the fact 
that, after having performed a simple exercise to assess the size
of such systematic error, we have found it to  be as large as 10~MeV. 
This is of the same order as the foreseen precision on $M_W$ at the end of the
LEP2 runs. However, we stress that our estimate should not 
be taken too literally, since our study lacks many of the necessary ingredients
of a complete simulation, such as detector effects, ISR, 
a more detailed background analysis (including all four-quark EW channels),
etc., all of which is  well beyond the scope of this paper. 

Nonetheless, it is evident that it is of crucial importance that a dedicated 
implementation based on  ${\cal O}(\alpha_s^2)$ QCD calculations
is used in the MC simulations of four-jet samples of the type
described in this paper. MC programs are now being upgraded
to incorporate such second-order dynamics and we urge the experimental 
collaborations to adopt them in their analyses.

\section*{Acknowledgements}

We are grateful to the UK PPARC for support, and to
the Theoretical Physics Department at Fermilab
for kind hospitality while part of this work was carried out. SM also
thanks the Department of Theoretical Physics in Lund for
hospitality, 
the Italian Institute of Culture 
`C.M. Lerici' for a grant (Ref. \#: Prot. I/B1 690) 
which supported his visit there, and T.~Sj\"ostrand
for useful conversations. Finally, we both thank Mike Seymour for carefully
reading the manuscript and for useful comments, as well as for 
pointing out a mistake in our analysis and for performing 
detailed comparisons between the MEs used in this paper and those implemented 
in the new HERWIG version.

\vfill
\newpage
\thispagestyle{empty}

\section*{Table Caption}

\begin{itemize}

\item[{[I]}] Fraction of four-jet events in various angular intervals
around the maxima of the $W^+W^-$ distributions, 
as predicted by the:
(i)   `${\cal O}(\alpha_s^2)$ MEs' ($q\bar q gg +q\bar q Q\bar Q$),
(ii)  `${\cal O}(\alpha_s)$ ME + PS' (HERWIG, {\tt IPROC=100}),
(iii) `${\cal O}(\alpha_s)$ ME + PS' (PYTHIA, {\tt ISUB=1}); and also,
for reference, the
`${\cal O}(\alpha_s^0)$ ME'  for \eeWWqqQQ.
All rates are obtained at parton level. The following cut has been implemented:
$|M_{ij}-M_{W}|<10$ GeV on at least two pairs of partons $(ij)$.
The CM energy is $\sqrt s=172$ GeV. 
Results are shown for the Durham algorithm, with $y_{\mathrm{cut}}=0.002$.
The first line refers to the case in which exactly $n=4$ jets are reconstructed
from the parton shower, while the second corresponds to $n\ge4$, 
with the additional jets in the latter eventually clustered into four.
(All $n$ final tracks are above cut-off in both cases.)
\end{itemize}

\vfill
\newpage
\thispagestyle{empty}

\section*{Figure Captions}

\begin{itemize}

\item[{[1]}] Feynman diagrams contributing in lowest order to QCD four-jet
production
in $e^+e^-$ annihilation: (a,b) double-gluon-bremsstrahlung,
(c) triple-gluon-vertex and (d) four-quark subprocesses.

\item[{[2]}] Differential distributions for the following variables:
(top-left) $\chi_{\mathrm{BZ}}$,
(top-right) $\Phi_{\mathrm{KSW}}^*$,
(bottom-left) $\cos\theta_{\mathrm{NR}}^*$ and
(bottom-right) $\cos\theta_{34}$ for
the following subprocesses: four-quark in QCD (solid), 
four-quark in a PS-like model
(dashed), triple-gluon in QCD (dotted),  triple-gluon in a PS-like model
(dot-dashed). The CM energy is $\sqrt s=M_Z$. All
distributions are normalised to unity.
Results are shown for the Durham algorithm, with $y_{\mathrm{cut}}=0.002$.

\item[{[3]}] The ratio between the ME and PYTHIA implementation (solid) 
of the differential distributions for the following variables:
(top-left) $\chi_{\mathrm{BZ}}$,
(top-right) $\Phi_{\mathrm{KSW}}^*$,
(bottom-left) $\cos\theta_{\mathrm{NR}}^*$ and
(bottom-right) $\cos\theta_{34}$. The default
spin correlation matrices of the parton cascade in the MC program have
been switched on. All distributions are at parton level. 
The CM energy is $\sqrt s=M_Z$. All
distributions are normalised to unity.
Results are shown for the Durham algorithm, with $y_{\mathrm{cut}}=0.008$.
Exactly four jets are required to be reconstructed from the PYTHIA cascade.
The corresponding ratio in case of HERWIG 6.1 is also shown (dashed): see
later on. 

\item[{[4]}] Differential distributions for the following variables:
(top-left) $\chi_{\mathrm{BZ}}$,
(top-right) $\Phi_{\mathrm{KSW}}^*$,
(bottom-left) $\cos\theta_{\mathrm{NR}}^*$ and
(bottom-right) $\cos\theta_{34}$ for
the following subprocesses: four-quark in QCD (solid), 
four-quark in a PS-like model
(dashed), triple-gluon in QCD (dotted),  triple-gluon in a PS-like model
(dot-dashed) and $W^+W^-$ (shaded). The CM energy is $\sqrt s=172$ GeV. All
distributions are normalised to unity. The following cut has been implemented:
$|M_{ij}-M_{W}|<10$ GeV on at least two pairs of partons $(ij)$.
Results are shown for the Durham algorithm, with $y_{\mathrm{cut}}=0.002$.

\item[{[5]}] Differential distributions for the following variables:
(top-left) $\chi_{\mathrm{BZ}}$,
(top-right) $\Phi_{\mathrm{KSW}}^*$,
(bottom-left) $\cos\theta_{\mathrm{NR}}^*$ and
(bottom-right) $\cos\theta_{34}$ in case of
`${\cal O}(\alpha_s^2)$ MEs' ($q\bar q gg +q\bar q Q\bar Q$, solid histogram),
`${\cal O}(\alpha_s)$ ME + PS' (HERWIG, dashed histogram) and
`${\cal O}(\alpha_s)$ ME + PS' (PYTHIA, dotted histogram). The default
spin correlation matrices of the parton cascade in the MC programs have
been switched on. For reference, we also
have included the same spectra as obtained from  the
`${\cal O}(\alpha_s^0)$ ME'  for \eeWWqqQQ\ (asterisk symbols).
All distributions are at parton level. The following cut has been implemented:
$|M_{ij}-M_{W}|<10$ GeV on at least two pairs of partons $(ij)$.
The CM energy is $\sqrt s=172$ GeV. Normalisation is to unity.
Results are shown for the Durham algorithm, with $y_{\mathrm{cut}}=0.002$.
(a) Exactly four jets are required to be reconstructed from the HERWIG and 
PYTHIA cascades.
(b) At least four jets are required to be reconstructed from the HERWIG and 
PYTHIA cascades, eventually forced into exactly four.

\item[{[6]}] Differential distributions for the following variables:
(top-left) $\chi_{\mathrm{BZ}}$,
(top-right) $\Phi_{\mathrm{KSW}}^*$,
(bottom-left) $\cos\theta_{\mathrm{NR}}^*$ and
(bottom-right) $\cos\theta_{34}$ in case of
`${\cal O}(\alpha_s^2)$ ME + PS + Hadronisation' (solid histogram) and
`${\cal O}(\alpha_s)$   ME + PS + Hadronisation' (dashed histogram), 
as obtained by using HERWIG. The default
spin correlation matrices of the parton cascade in the two versions of
the MC program have
been switched on. The following cut has been implemented:
$|M_{ij}-M_{W}|<10$ GeV on at least two pairs of jets $(ij)$, as defined
at the hadron level.
The CM energy is $\sqrt s=172$ GeV. Normalisation is to unity.
Results are shown for the Durham algorithm, with $y_{\mathrm{cut}}=0.002$.
Exactly four jets are required to be reconstructed.

\item[{[7]}] Differential distributions in the average $W^\pm$-mass
(as defined in the text) for the $W^+W^-$ signal (solid curve),
the ${\cal O}(\alpha_s)$   (dashed curve) and 
the ${\cal O}(\alpha_s^2)$ (dotted curve) backgrounds
(from HERWIG).
Other than the constraint
$|M_{ij}-M_{W}|<10$ GeV on at least two pairs of jets $(ij)$, as defined
at the hadron level, the angular cuts of eq.~(\ref{cuts}) have also
been implemented.
The CM energy is $\sqrt s=172$ GeV. Normalisation is to unity.
Results are shown for the Durham algorithm, with $y_{\mathrm{cut}}=0.002$.
Exactly four jets are required to be reconstructed.

\item[{[8]}] Differential distributions (top) in the average $W^\pm$-mass
(as defined in the text) for the sum of the $W^+W^-$ signal 
and the ${\cal O}(\alpha_s)$(${\cal O}(\alpha_s^2)$) background (from HERWIG):
 dashed(dotted) curve.
Other than the constraint
$|M_{ij}-M_{W}|<10$ GeV on at least two pairs of jets $(ij)$, as defined
at the hadron level, the angular cuts of eq.~(\ref{cuts}) have also
been implemented.
The CM energy is $\sqrt s=172$ GeV. Normalisation is to the total rates.
Results are shown for the Durham algorithm, with $y_{\mathrm{cut}}=0.002$.
Exactly four jets are required to be reconstructed.
The bottom plot shows the ratio between the above two spectra.

\end{itemize}

\vfill
\clearpage
\thispagestyle{empty}

\begin{table}[htb]
\begin{center}
\begin{tabular}{|c||c||c|c|c|}

\hline
interval & $W^+W^-$ ME & QCD MEs & HERWIG & PYTHIA \\ 
\hline\hline

\multicolumn{5}{|c|}
{$\chi_{\mathrm{BZ}}$}
\\ \hline
$ (75,180)^\circ$ & $0.849$ & $0.666$ & $0.608$ & $0.584$ \\
                  & $$      & $$      & $0.596$ & $0.574$ \\
$(100,180)^\circ$ & $0.819$ & $0.585$ & $0.529$ & $0.529$ \\
                  & $$      & $$      & $0.492$ & $0.495$ \\
$(125,180)^\circ$ & $0.732$ & $0.460$ & $0.412$ & $0.436$ \\
                  & $$      & $$      & $0.361$ & $0.388$ \\
\hline\hline

\multicolumn{5}{|c|}
{$\Phi_{\mathrm{KSW}}^*$}
\\ \hline
$ (75,180)^\circ$ & $0.896$ & $0.748$ & $0.708$ & $0.718$ \\
                  & $$      & $$      & $0.670$ & $0.684$ \\
$(100,180)^\circ$ & $0.827$ & $0.641$ & $0.601$ & $0.614$ \\
                  & $$      & $$      & $0.554$ & $0.573$ \\
$(125,180)^\circ$ & $0.666$ & $0.485$ & $0.446$ & $0.468$ \\
                  & $$      & $$      & $0.399$ & $0.424$ \\
\hline\hline

\multicolumn{5}{|c|}
{$\theta_{\mathrm{NR}}^*$}
\\ \hline
$   (0,1)$ & $0.829$ & $0.549$ & $0.493$ & $0.493$ \\
           & $$      & $$      & $0.476$ & $0.471$ \\
$(0.25,1)$ & $0.803$ & $0.400$ & $0.366$ & $0.375$ \\
           & $$      & $$      & $0.355$ & $0.359$ \\
$ (0.5,1)$ & $0.747$ & $0.244$ & $0.239$ & $0.246$ \\
           & $$      & $$      & $0.238$ & $0.242$ \\
\hline\hline

\multicolumn{5}{|c|}
{$\theta_{34}$}
\\ \hline
$(-0.8,0.8)$ & $0.785$ & $0.635$ & $0.565$ & $0.529$ \\
             & $$      & $$      & $0.649$ & $0.603$ \\
$(-0.6,0.6)$ & $0.583$ & $0.422$ & $0.356$ & $0.324$ \\
             & $$      & $$      & $0.442$ & $0.400$ \\
$(-0.4,0.4)$ & $0.384$ & $0.266$ & $0.219$ & $0.199$ \\
             & $$      & $$      & $0.283$ & $0.254$ \\
\hline\hline

\multicolumn{5}{|c|}
{$e^+e^-\ar 4$ jets at LEP2}
\\ \hline\hline

\multicolumn{5}{|c|}
{$|M_{ij}-M_{W}|<10$ GeV}
\\ \hline
\end{tabular}
\end{center}
\vskip1.0cm
\centerline{\bf\Large Tab.~I}
\end{table}

\vfill
\clearpage
\thispagestyle{empty}

\begin{figure}[!t]
~\epsfig{file=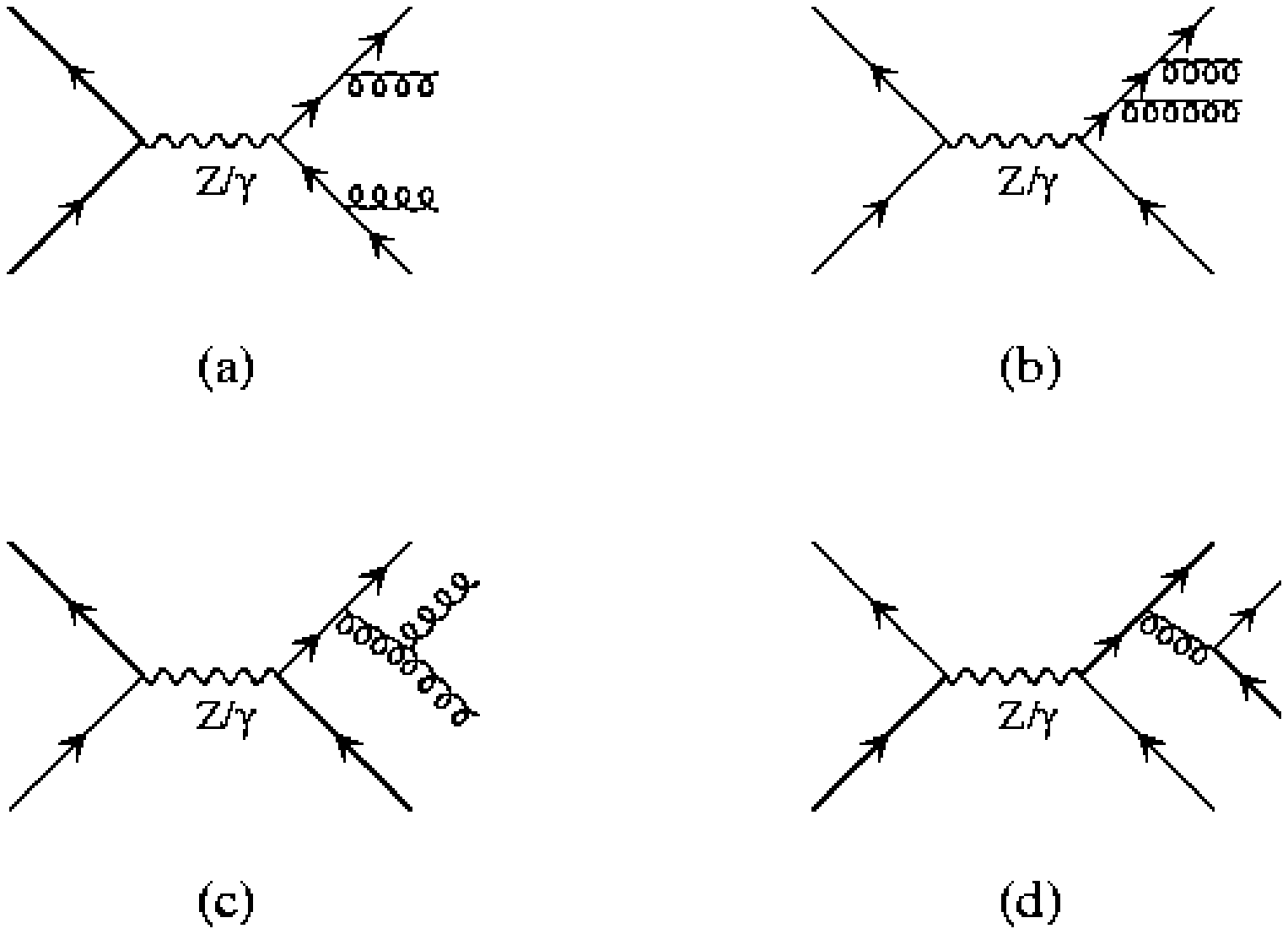,height=22cm}

\centerline{\large\bf Fig.~1}
\end{figure}
\stepcounter{figure}
\vfill
\clearpage
\thispagestyle{empty}

\begin{figure}[!t]
~\epsfig{file=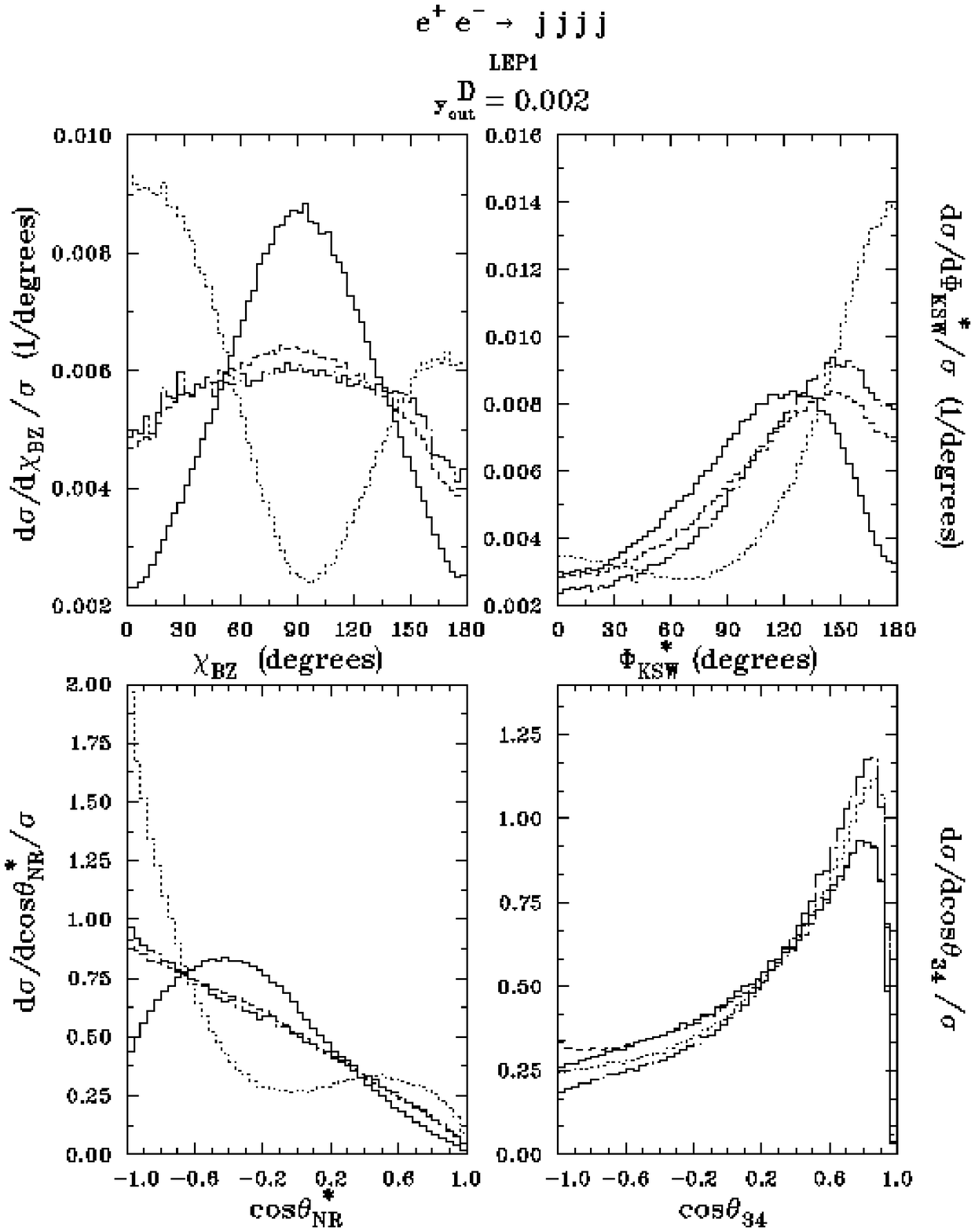,height=22cm}

\centerline{\large\bf Fig.~2}
\end{figure}
\stepcounter{figure}
\vfill
\clearpage
\thispagestyle{empty}

\begin{figure}[t!]
\begin{minipage}[b]{.5\linewidth}
\centering\epsfig{file=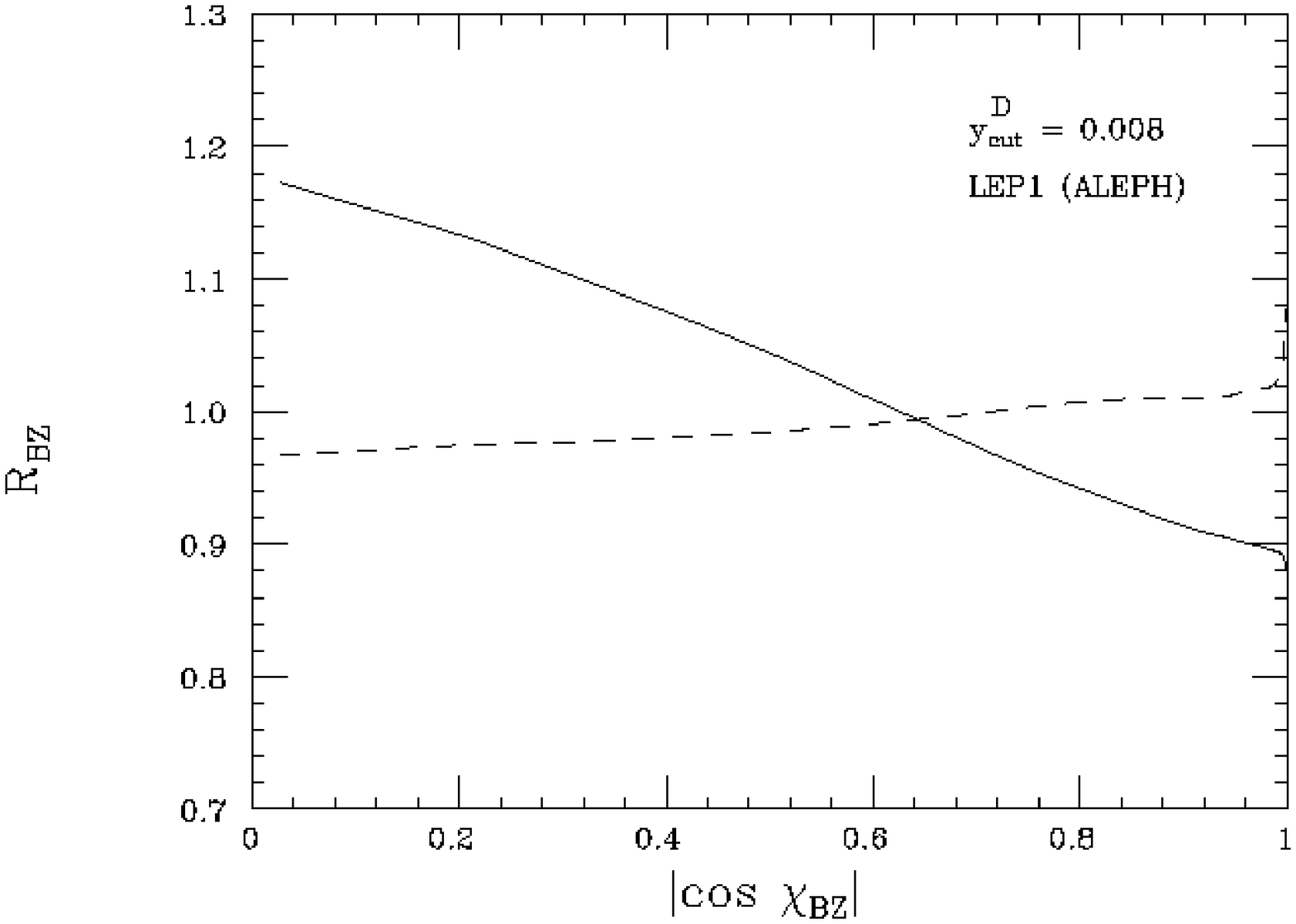,angle=90,height=6cm,width=\linewidth}
\end{minipage}\hfill
\begin{minipage}[b]{.5\linewidth}
\centering\epsfig{file=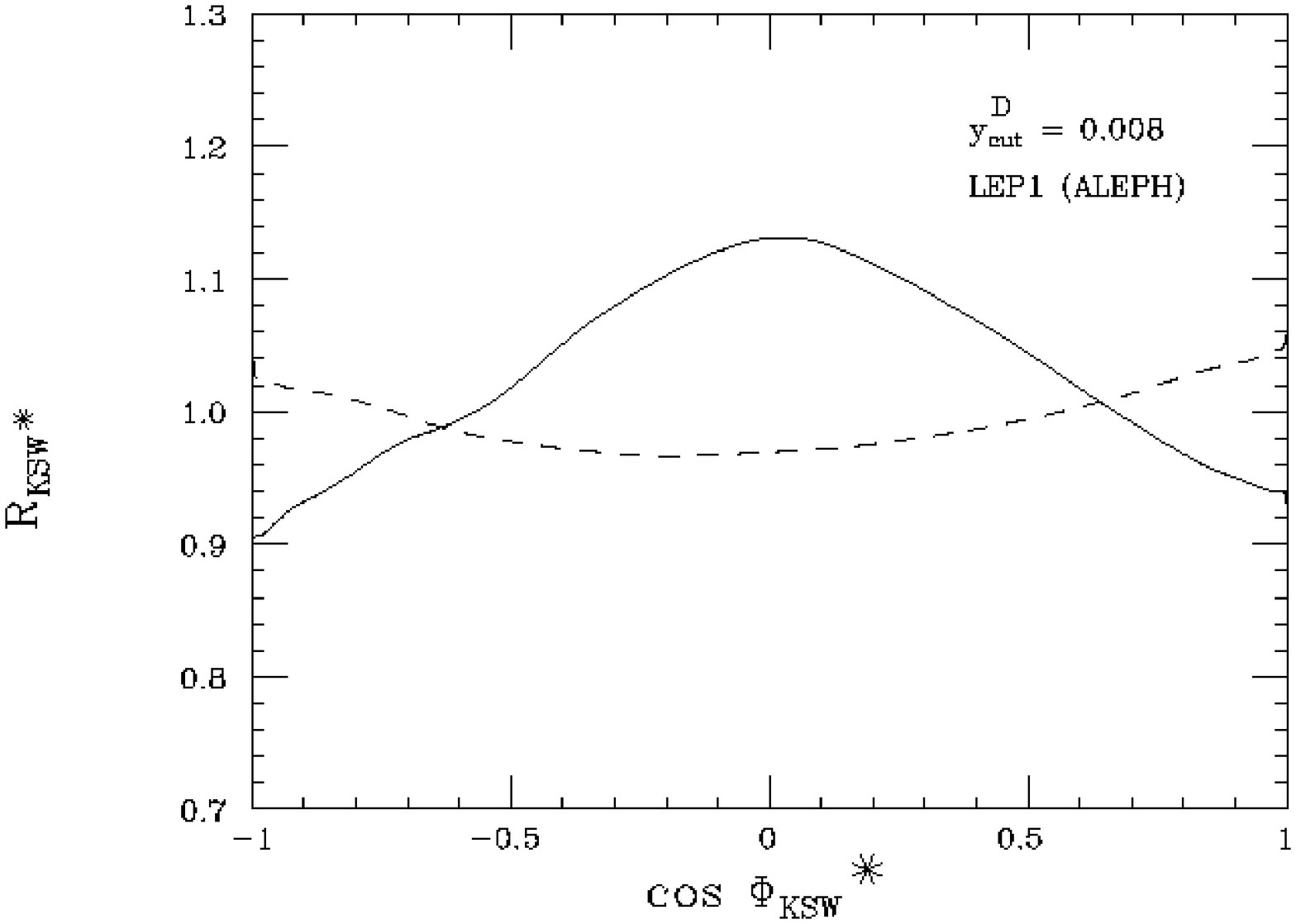,angle=90,height=6cm,width=\linewidth}
\end{minipage}\hfill
\begin{minipage}[b]{.5\linewidth}
\centering\epsfig{file=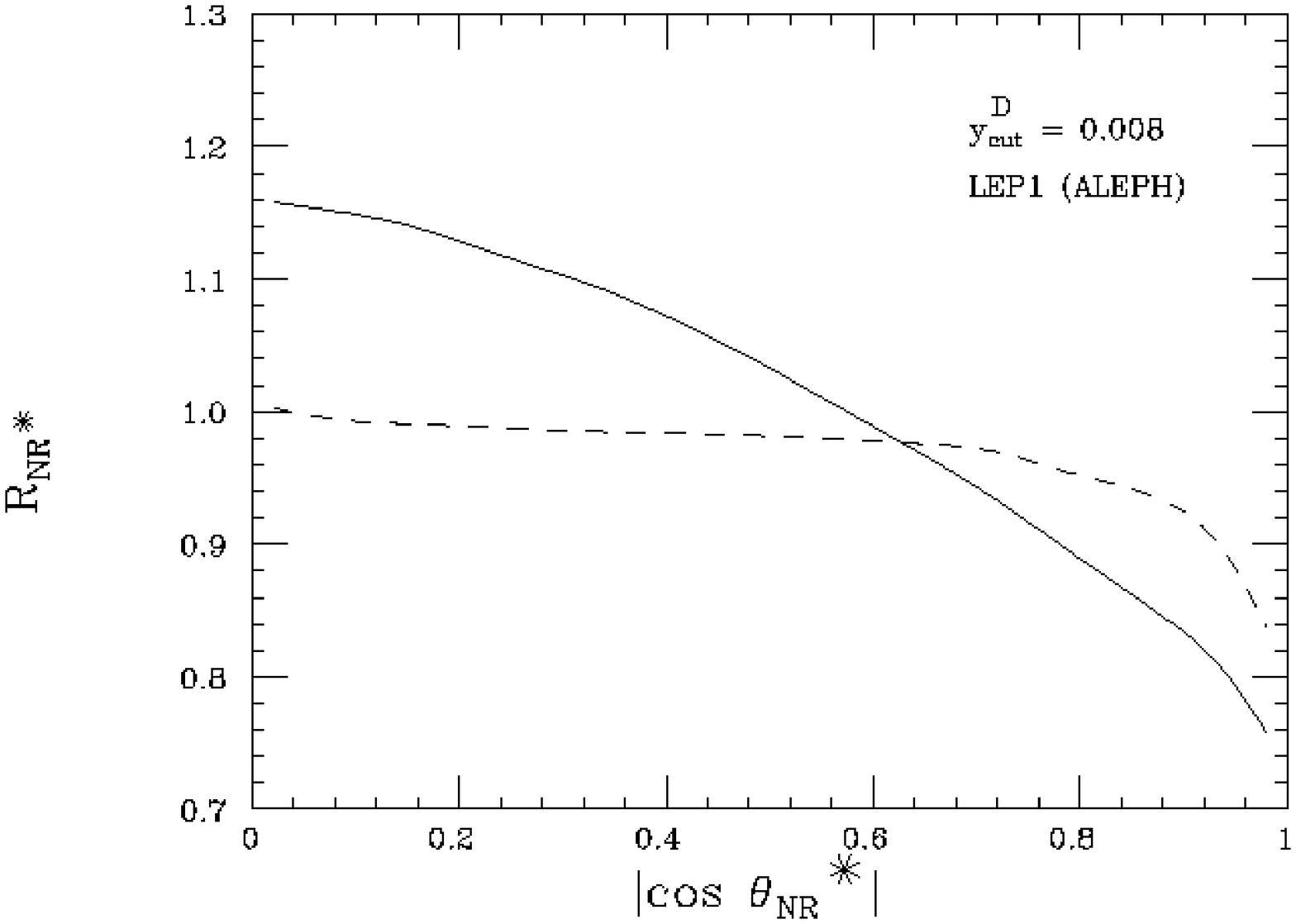,angle=90,height=6cm,width=\linewidth}
\end{minipage}\hfill
\begin{minipage}[b]{.5\linewidth}
\centering\epsfig{file=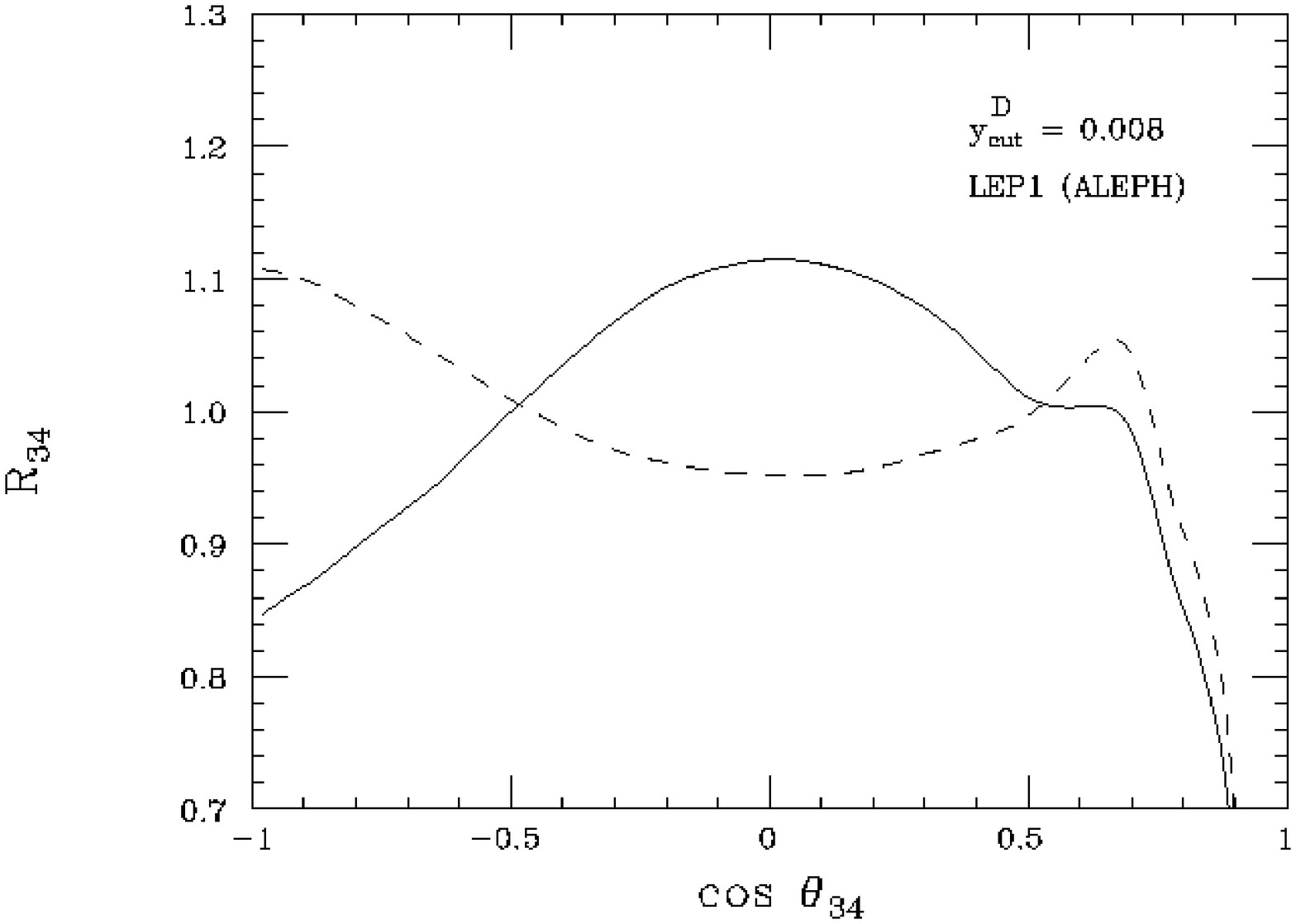,angle=90,height=6cm,width=\linewidth}
\end{minipage}\hfil

\centerline{\large\bf Fig.~3}
\end{figure}
\stepcounter{figure}
\vfill
\clearpage
\thispagestyle{empty}

\begin{figure}[t]
~\epsfig{file=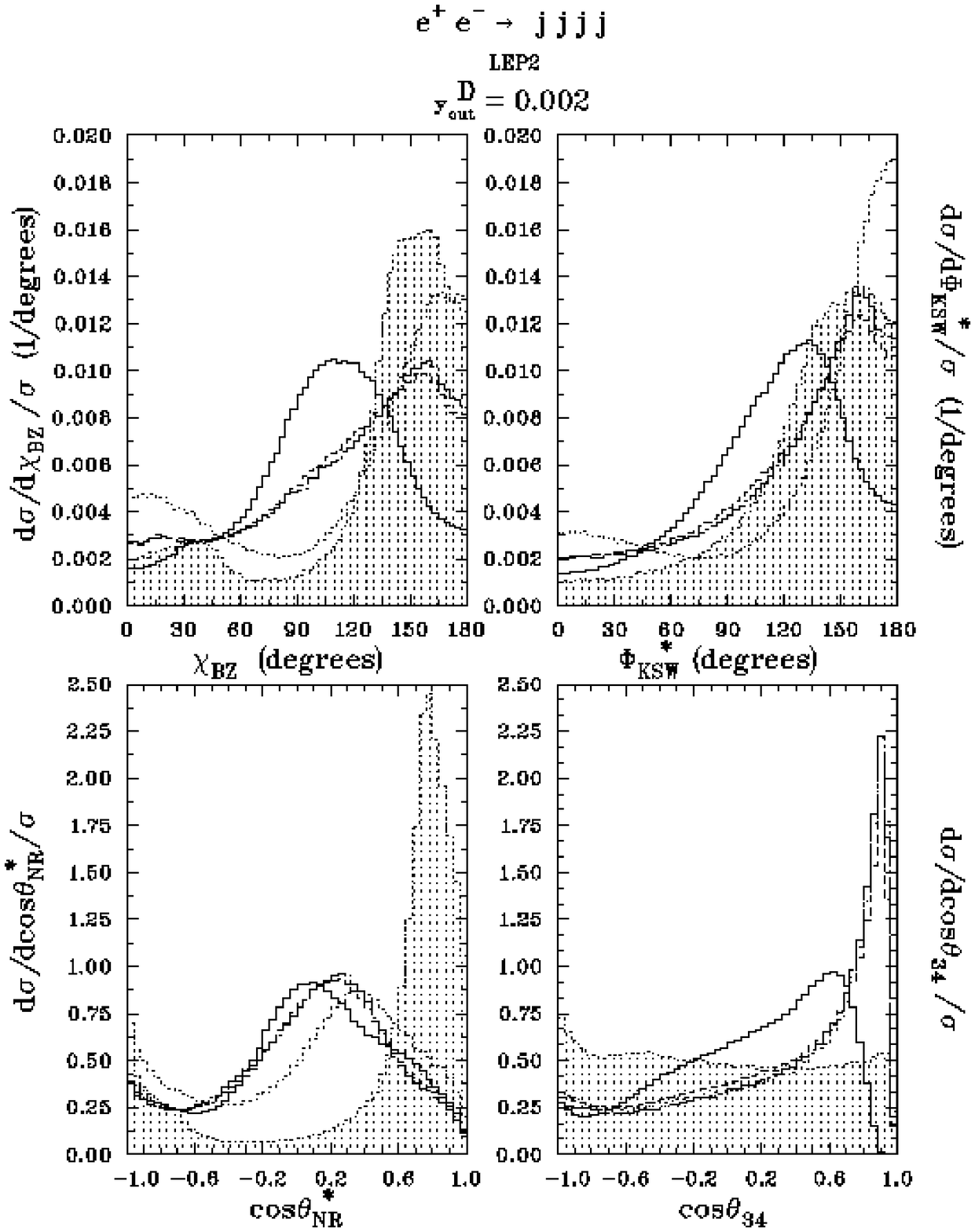,height=22cm}

\centerline{\large\bf Fig.~4}
\end{figure}
\stepcounter{figure}
\vfill
\clearpage
\thispagestyle{empty}

\begin{figure}[t]
~\epsfig{file=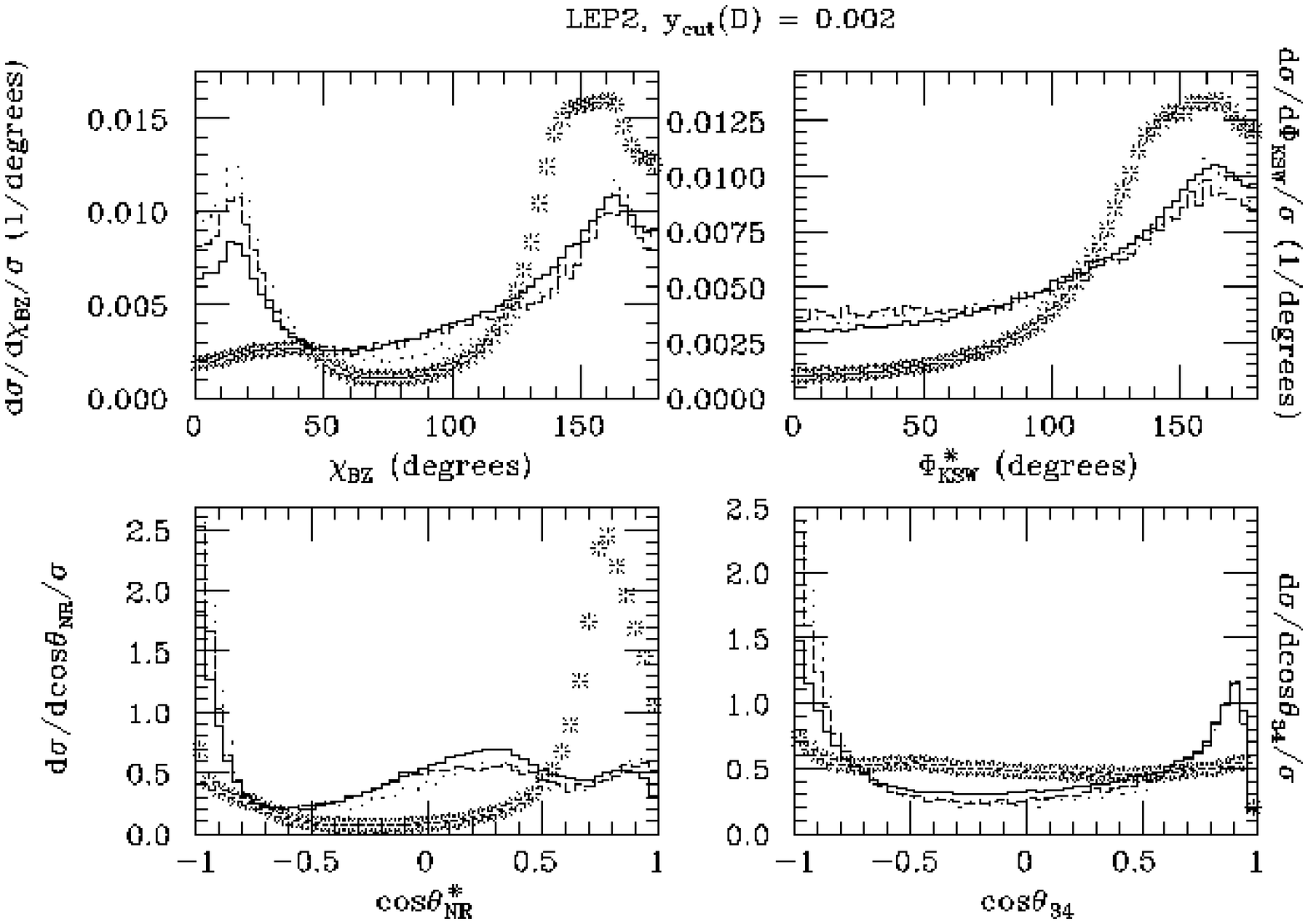,height=16cm,angle=90}

\centerline{\large\bf Fig.~5a}
\end{figure}
\stepcounter{figure}

\vfill
\clearpage
\thispagestyle{empty}

\begin{figure}[t]
~\epsfig{file=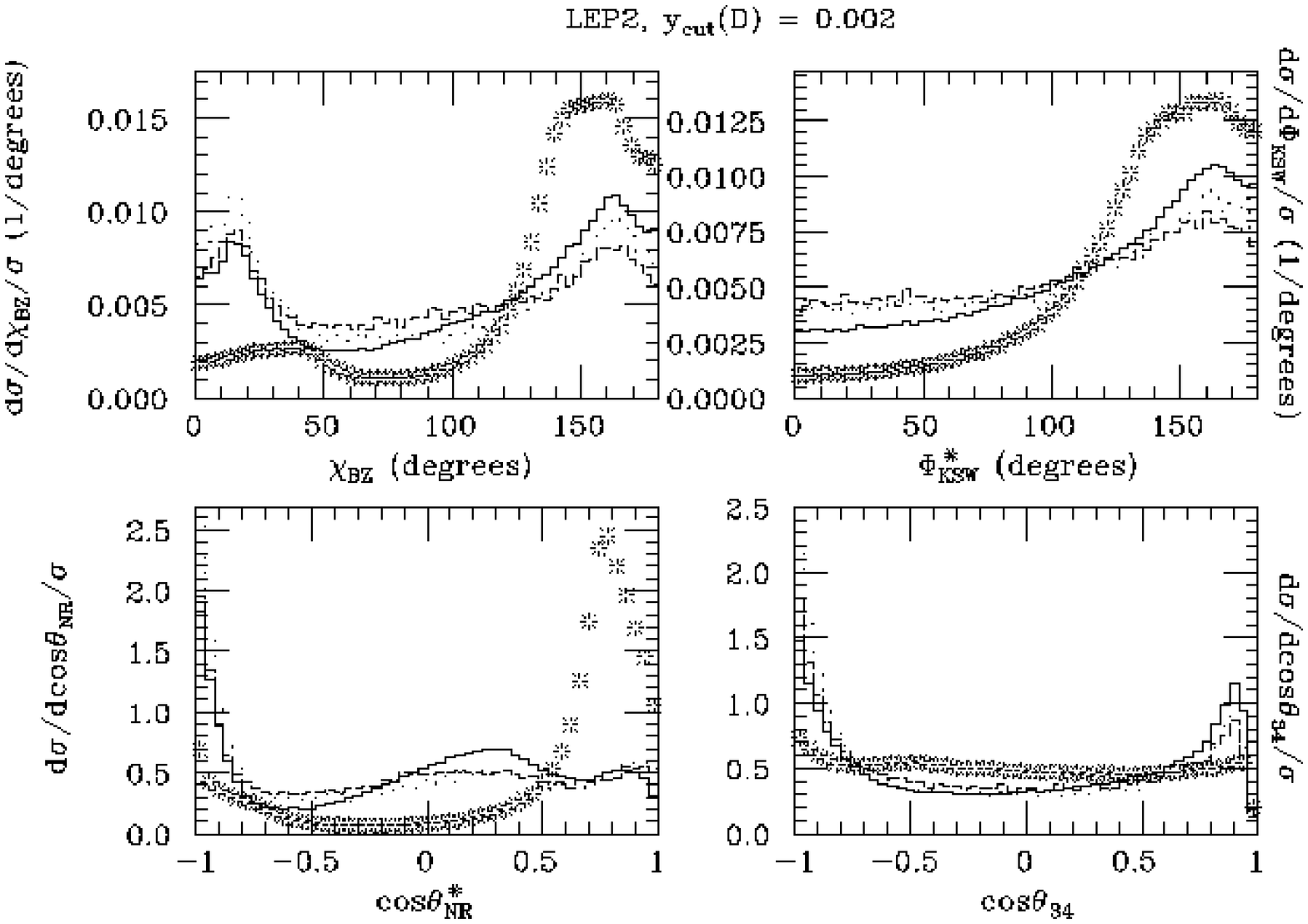,height=16cm,angle=90}

\centerline{\large\bf Fig.~5b}
\end{figure}
\stepcounter{figure}

\begin{figure}[t]
~\epsfig{file=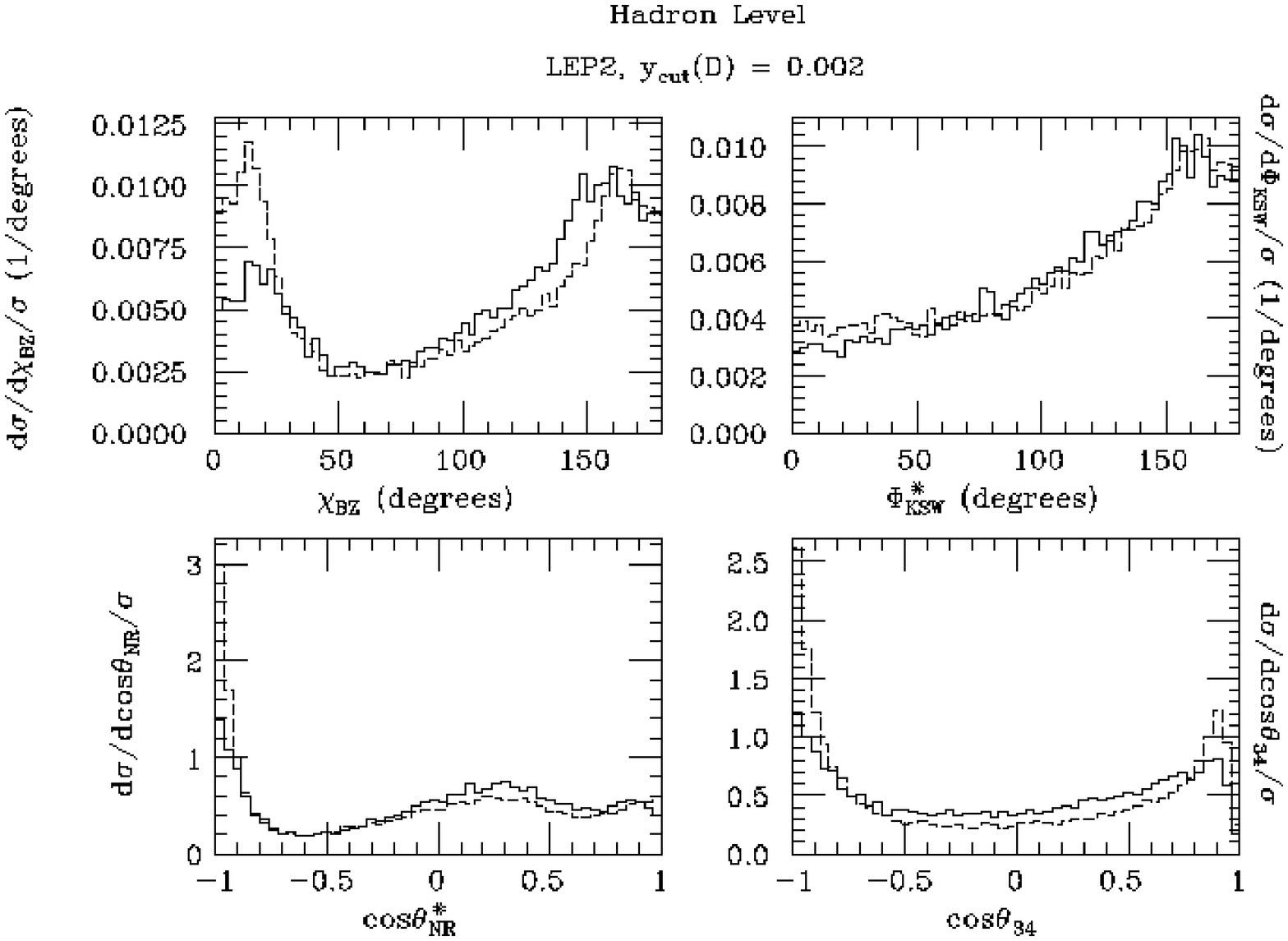,height=16cm,angle=90}

\centerline{\large\bf Fig.~6}
\end{figure}
\stepcounter{figure}

\begin{figure}[t]
~\epsfig{file=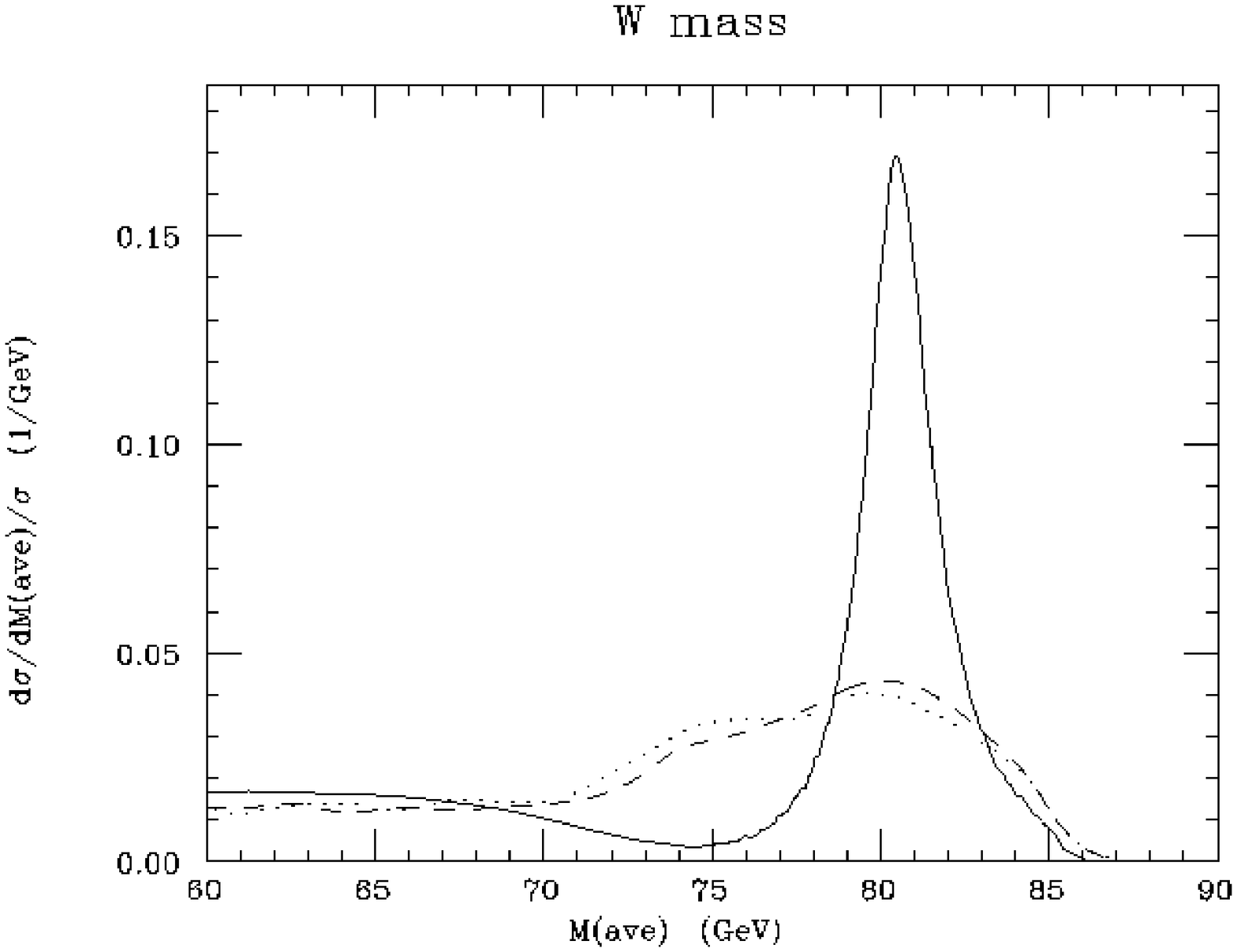,height=16cm,angle=90}

\centerline{\large\bf Fig.~7}
\end{figure}
\stepcounter{figure}

\begin{figure}[t]
~\epsfig{file=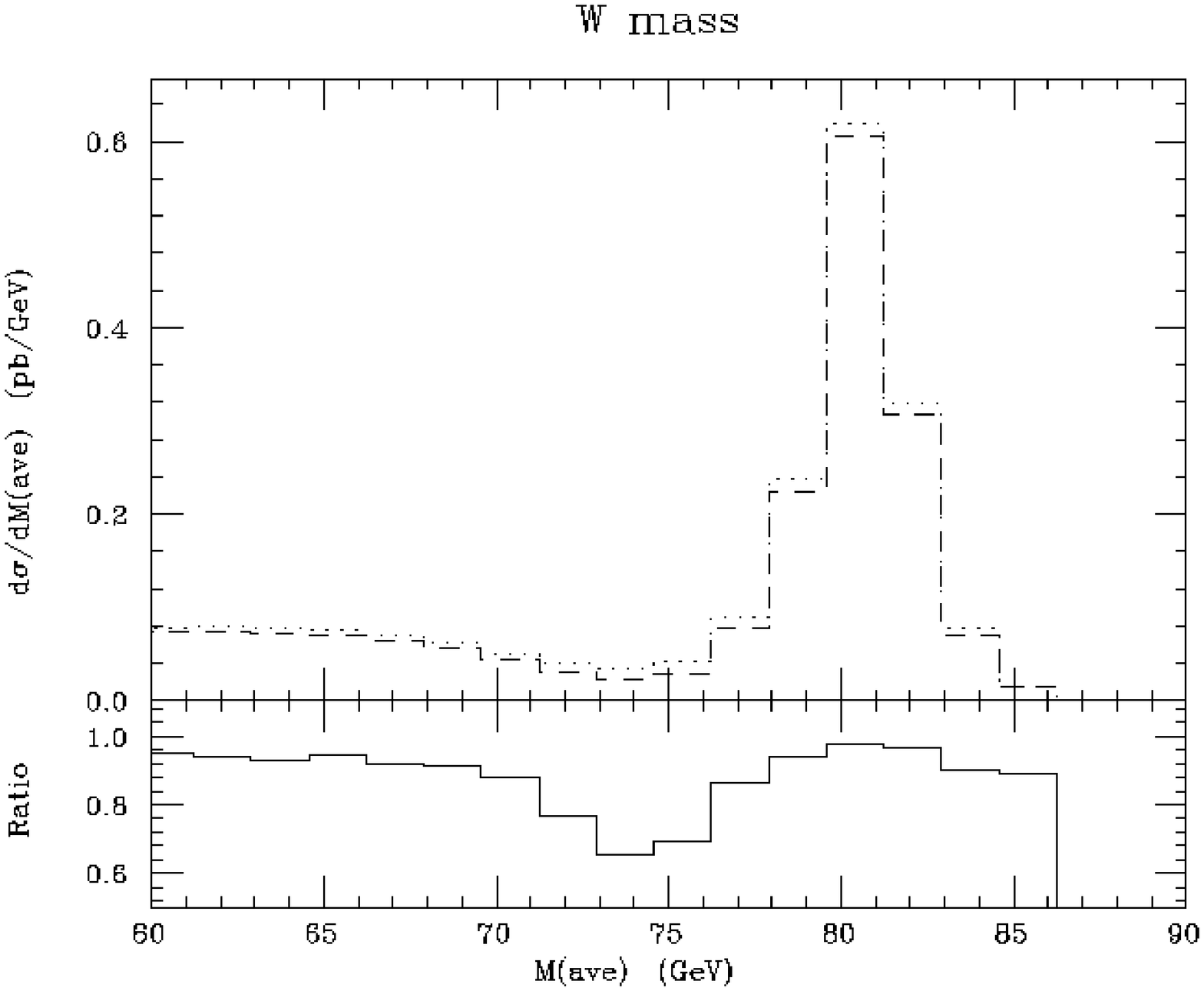,height=16cm,angle=90}

\centerline{\large\bf Fig.~8}
\end{figure}
\stepcounter{figure}

\end{document}